\documentclass[conference,9pt]{IEEEtran}

\usepackage{algpseudocode}
\algtext*{EndWhile}
\algtext*{EndIf}
\algtext*{EndFor}
\algtext*{EndProcedure}
\usepackage{amsmath}
\usepackage{amssymb}
\usepackage{array}
\usepackage{algorithm}
\usepackage{bm}
\usepackage[bookmarks=false, hidelinks]{hyperref}
\usepackage[noadjust]{cite}
\usepackage[font=small]{caption}
\usepackage{color}
\usepackage{comment}
\usepackage{dblfloatfix}
\usepackage[inline]{enumitem}
\usepackage{epsfig}
\usepackage{etoolbox}
\usepackage{fancybox}
\usepackage{float}
\usepackage{fixltx2e}
\usepackage{graphicx}
\usepackage{listings}
\usepackage{mathrsfs}
\usepackage{multirow}
\usepackage{multicol}
\usepackage{pifont}
\usepackage{placeins}
\usepackage{rotating}
\usepackage{setspace}
\usepackage{soul}
\usepackage[hang, tight]{subfigure}
\usepackage{threeparttable}
\usepackage{times}
\usepackage{url}
\usepackage{upgreek}
\usepackage{verbatim}
\usepackage{wrapfig}
\usepackage[usenames,dvipsnames]{xcolor}
\usepackage{authblk}
\usepackage[]{siunitx}
\usepackage[super]{nth}
\usepackage{tabularx}
\usepackage{supertabular}
\usepackage{tikz}
\usepackage{bbm}
\usepackage{blindtext}

\usepackage{geometry}
\graphicspath{{./_fig/}}

\usepackage{bbding}

\makeatletter
\renewcommand\footnoterule{%
  \kern-3\p@
  \hrule\@width0.4\columnwidth
  \kern2.6\p@}
  \makeatother

\usepackage{tikz}
\usepackage{xcolor}

\usepackage[hang,flushmargin]{footmisc}

\begin{document}

%\title{WhitePower: Generalizable White-Box Architecture-Level Power Modeling with Intelligent Parameter Calibration}

%\title{READY: A \underline{Re}liable, Interpret\underline{a}ble, and Han\underline{dy} Architectural Power Model Based on Analytical Framework}

\title{\vspace{-.2in}ReadyPower: A \underline{Re}liable, Interpret\underline{a}ble, and Han\underline{dy} Architectural \underline{Power} Model Based on Analytical Framework\vspace{-.15in}}

%Inter- and Intra-Power Group Decoupling}
% Automating Data-Friendly Architecture-Level Power Modeling by Decoupling Across and Within Power Groups 

\author[]{ \fontsize{11}{11}\selectfont Qijun Zhang, Shang Liu, Yao Lu, Mengming Li, Zhiyao Xie\textsuperscript{*}\vspace{-6pt}}

\affil[]{\fontsize{10}{10}\selectfont Hong Kong University of Science and Technology\vspace{-7pt}}

\affil[]{	\{qzhangcs, sliudx, yludf, mengming.li\}@connect.ust.hk, eezhiyao@ust.hk\vspace{-7pt}}
\maketitle

\begingroup\renewcommand\thefootnote{*}
\footnotetext{Corresponding Author}
\endgroup

\begin{abstract}

Power is a primary objective in modern processor design, requiring accurate yet efficient power modeling techniques. Architecture-level power models are necessary for early power optimization and design space exploration. However, classical analytical architecture-level power models (e.g., McPAT) suffer from significant inaccuracies. Emerging machine learning (ML)-based power models, despite their superior accuracy in research papers, are not widely adopted in the industry. In this work, we point out three inherent limitations of ML-based power models: unreliability, limited interpretability, and difficulty in usage. 

This work proposes a new analytical power modeling framework named ReadyPower, which is ready-for-use by being reliable, interpretable, and handy. We observe that the root cause of the low accuracy of classical analytical power models is the discrepancies between the real processor implementation and the processor’s analytical model. To bridge the discrepancies, we introduce architecture-level, implementation-level, and technology-level parameters into the widely adopted McPAT analytical model to build ReadyPower. The parameters at three different levels are decided in different ways. In our experiment, averaged across different training scenarios, ReadyPower achieves $>20\%$ lower mean absolute percentage error (MAPE) and $>0.2$ higher correlation coefficient $R$ compared with the ML-based baselines, on both BOOM and XiangShan CPU architectures.

% state-of-the-art solution, McPAT-Calib

% For each processor architecture, ReadyPower decides

% In our experiments, across different training cases, ReadyPower can achieve a low average mean absolute percentage error (MAPE) of 9.26\% and 14.51\% and a high average correlation $R$ of 0.96 and 0.91, even with only three available configurations, on BOOM and XiangShan CPU. This is 20.76\% and 29.10\% lower in MAPE and 0.25 and 0.24 higher in correlation $R$ compared with the enhanced version of McPAT-Calib, the representative ML-based power model.

\end{abstract}

\section{Introduction}

Power is a primary design objective in modern processor design, requiring accurate yet efficient power modeling techniques. 
The standard power estimation process requires the full VLSI design flow—requiring Register-Transfer Level (RTL) implementation, followed by RTL simulation, logic synthesis, and gate-level power analysis with EDA tools~\cite{vcs,ptpx,design-compilier}. While this process delivers high accuracy, it incurs substantial manpower and runtime, making it impractical for rapid design iterations. As a result, architecture-level power models are in high demand for early power optimization and design space exploration. There are two types of architecture-level power models: 1) conventional analytical models, and 2) recent machine learning (ML)-based models, as we introduce below.

\textbf{Analytical Power Model:} Conventional architecture-level power models are analytical models, such as McPAT~\cite{li2009mcpat} and Wattch~\cite{brooks2000wattch}. These models require engineers to meticulously characterize each microarchitectural component within a target processor~\cite{xi2015quantifying}. As a result, they typically suffer from significant inaccuracies when applied to new architectures, as indicated in many existing studies~\cite{xi2015quantifying}. 

% constrained to specific processors and 

\textbf{ML-based Power Model:} In recent years, machine learning (ML)-based architecture-level power models emerged to address the accuracy limitations of traditional analytical models. These data-driven models construct either black-box~\cite{zhai2022mcpat} or gray-box~\cite{zhang2023panda} representations of power consumption. The models are trained with known processor design configurations with ground-truth power values. When trained with appropriate data, these ML-based models have demonstrated high accuracy compared to analytical models.

\textbf{Rethinking: why are ML-based models not widely adopted?} Despite the clearly superior accuracy of recent ML-based power models~\cite{zhai2022mcpat,lee2015powertrain,zhang2023panda}, unexpectedly, we notice that these new ML-based methods are not widely adopted in the industry. 
Instead, architects continue to utilize analytical models such as McPAT, although substantial engineering effort is required to fix the inaccuracies. This observation motivates this work, which starts with rethinking the problem of existing research on ML-based power models. We point out three inherent limitations of ML-based power models below.

%The limitation of PANDA~\cite{zhang2023panda} and FirePower~\cite{zhang2025firepower} is obvious: they are not automatic, where PANDA requires significant engineering effort to build the resource function and FirePower relies on existing known architecture.
%Besides the automation, our analysis reveals two fundamental limitations hindering the widespread deployment of even automatic ML-based approaches such as McPAT-Calib~\cite{zhai2022mcpat}:

%They exhibit significant accuracy degradation when testing data distribution deviates from the training data, particularly under limited training data scenarios. 

\begin{figure}[!t]
\centering
\vspace{-.15in}
\includegraphics[width=0.4\textwidth]{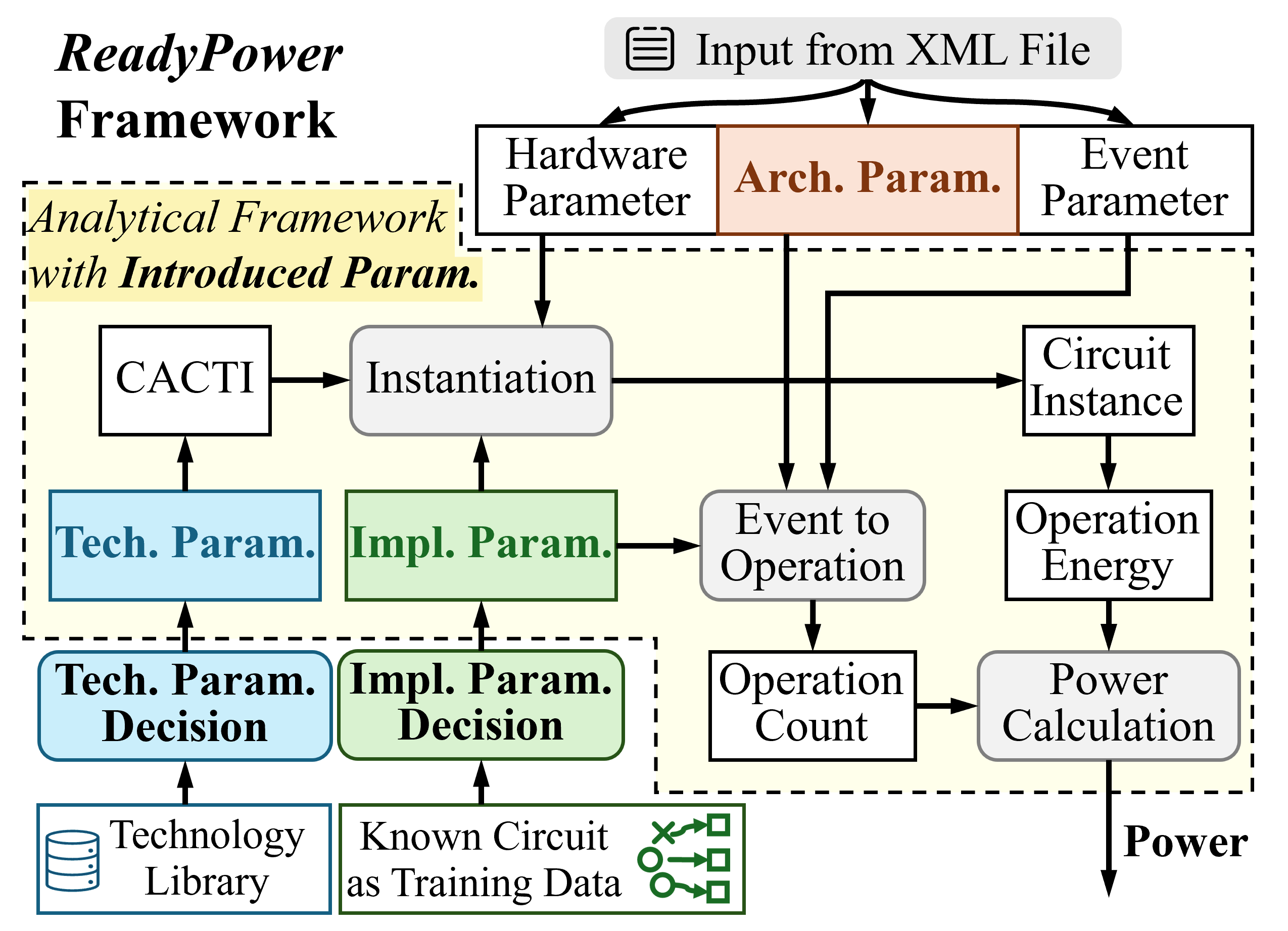}
\vspace{-.05in}
\caption{The ReadyPower framework. ReadyPower fixes the discrepancies between real processor implementation and classical power models. It introduces new architecturally interpretable parameters at the architecture, implementation, and technology levels within the analytical framework.}
\vspace{-.2in}
\label{framework}
\end{figure}

\begin{enumerate}
\item \textbf{Unreliability:} ML models are inherently weak at \emph{extrapolation}, since ML models are performing \emph{interpolation}\footnote{Interpolation is performing inference within the range (i.e., distribution) of the known training data, while extrapolation goes beyond the range.} based on training data. As a result, ML models can become extremely unreliable when the testing data falls \emph{out of the distribution} of training data. 
For instance, models trained only on small processor configurations will be highly inaccurate when applied to large configurations, and vice versa\footnote{All existing ML-based power modeling works~\cite{zhai2022mcpat,lee2015powertrain,zhang2023panda} evade this limitation by using very similar training and testing configurations in the experiment.}. 
\item \textbf{Limited Interpretability:} 
ML-based power models are inherently black-box, preventing the interpretation of their power calculation mechanism. The parameters (i.e., weights) of ML-based power models are calculated by directly fitting training data, without representing any realistic physical information. 
%. The model parameters do not have physical meanings. %represent actual . 
%\st{While these models effectively capture feature-label correlations, the inability to trace detailed power consumption computations restricts their utility for architectural analysis and targeted optimization.} \yao{[such `inability is not clear']}
%%
\item \textbf{Difficulty for Usage:} % The last limitation is about the deployment and usage. 
Existing ML-based power models are mostly research works and involve complex training stages before being ready for use. They do not provide a consistent user interface that can be easily used by designers or incorporated into the existing design environment. \looseness=-1
%designers can easily use. 
\end{enumerate}

Besides these three inherent limitations, some recent ML-based power models~\cite{zhang2023panda} further require significant designer expertise to customize the model towards each new target processor architecture, setting an additional barrier to the adoption of ML-based models.

\textbf{ReadyPower:} In this work, we propose a novel architecture-level power model named ReadyPower, addressing the aforementioned challenges. As its name suggests, ReadyPower stands for a ready-for-use power model by being reliable, interpretable, and handy\footnote{Our proposed ReadyPower framework has been fully open-sourced at https://github.com/hkust-zhiyao/ReadyPower. It adopts the same user interface as McPAT, ready for designers to directly use it.}. 
Different from all existing ML-based power models, as shown in Fig.~\ref{framework}, our solution adopts the original framework and interface of a classical analytical model (i.e., McPAT~\cite{li2009mcpat}) for easy deployment and usage. Specifically, we introduce new parameters internally inside McPAT's analytical framework, and decide the parameter values based on a few known designs (i.e., training data).

ReadyPower is designed based on a key observation: the root cause of inaccurate analytical models is the \emph{discrepancies between the real processor implementation and the processor’s analytical model}, and the discrepancies exist in all three major levels (architecture, implementation, and technology levels). We will further elaborate on this observation in Sec.~\ref{sec:discrepancy}. As Fig.~\ref{framework} shows, ReadyPower bridges the discrepancies by introducing \emph{new parameters} at all three levels into the analytical framework:  
1) Architecture-level new parameters capture important microarchitecture design decisions. These parameters can be directly set by designers as input, since they directly reflect architecture decisions. 2) Implementation-level new parameters capture power characteristics that depend on RTL and downstream design implementation. These parameters' values are decided based on training data. The gradient of each new parameter is approximated, and then gradient descent is performed for training. 3) Technology-level new parameters reflect the process technology. These parameters are design-agnostic, decided solely based on technology libraries.

%\yao{[ML model unreliable (testing data out of range), we set a reasonable range]}  

ReadyPower solves the three inherent limitations of ML-based power models:  
1) \emph{Reliable:} 
ReadyPower maintains consistent high accuracy even when testing data falls out of the training data distribution. 
Instead of directly fitting the ultimate total power value, ReadyPower fixes discrepancies at each level within the analytical framework. Moreover, it set a reasonable range for all introduced parameters. This framework prevents the unexpected large variations in complex ML models' predictions.   
2) \emph{Interpretable:} 
Unlike black-box alternatives, each new parameter introduced in ReadyPower corresponds to an explicit physical parameter in the processors. Each step of the power calculation is also explicitly defined and architecturally interpretable. 
3) \emph{Ready-to-use:} 
ReadyPower adopts the same user interface as the McPAT~\cite{li2009mcpat}. Therefore, when deployed in the industry, it can be easily integrated into the existing frameworks with McPAT as a component, without any modification to other components.
\looseness=-1

\section{Background}
% \yao{[why start with analytical]} 
% Our proposed ReadyPower is built upon the analytical architecture-level power model. In this section, we introduce the general formulation of the architecture-level power model and the basic power estimation approach in the McPAT~\cite{li2009mcpat}.

\subsection{Architecture-Level Power Model}
Generally, architecture-level power model takes hardware parameters and event parameters as input to estimate processor power consumption when running a workload. The hardware parameters $H$ represent processor configurations, such as FetchWidth and DCacheWay. Event parameters $E$ are architecture-level events when a processor executes a workload, such as the number of branch mispredictions and the number of data cache misses. Event parameters are collected from architecture-level performance simulators, such as gem5~\cite{binkert2011gem5}, by simulating a workload on a specific processor configuration.
\looseness=-1

% We take the McPAT~\cite{li2009mcpat} as our example to explain the basic power estimation approach in the analytical power model. Building on the power estimation flow of the analytical power model, we analyze the limitations %based on our understanding of a processor design flow
% and summarize these limitations into a three-level design-flow-aware hierarchy.

% \subsection{Power Estimation of ML-based Model}

In recent years, ML solutions are widely adopted in EDA applications~\cite{xie2018routenet,zhang2023panda,zhang2025firepower,zhang2025autopower,zhang2024architecture,lu2024unleashing,li2025atlas,fang2024transferable,fang2023masterrtl,xie2021apollo,xie2022deep}. Existing architecture-level power models include the analytical model and the ML-based model~\cite{zhang2023panda}. Analytical power models~\cite{li2009mcpat,brooks2000wattch} derive individual models for each component based on the background knowledge of model developers. It first models the energy consumption of each event based on hardware parameters $H$, and then calculates the final power with event parameters $E$.

ML-based power models~\cite{zhai2022mcpat,zhai2023microarchitecture,zhang2023panda} directly learn the correlation between the input features (i.e., hardware parameters $H$ and event parameters $E$) and the final power consumption with an ML model automatically. However, as mentioned in the Introduction, ML solutions can be unreliable, uninterpretable, and difficult to use.

% For other variations, ASP-DAC'23\cite{zhai2023microarchitecture} can not be transferred to unknown configurations. PANDA~\cite{zhang2023panda} relies on human-defined resource functions, leading to an unautomated solution that hinders ease of use. FirePower~\cite{zhang2025firepower} requires a well-developed existing architecture and is not a standalone solution for processor power modeling.     

\subsection{Analytical Model Framework}

% \begin{figure}[!t]
% \centering
% %\vspace{-.1in}
% \includegraphics[width=0.48\textwidth]{_fig/mcpat.png}
% %\vspace{-.1in}
% \caption{Power estimation of analytical power model.}
% %\vspace{-.2in}
% \label{mcpat}
% \end{figure}

We introduce the representative analytical power model, McPAT~\cite{li2009mcpat}, whose analytical framework is adopted in ReadyPower. 
McPAT takes an XML file as its input. The XML file includes the hardware parameters $H$ and event parameters $E$.  
%The hardware parameters may describe the component scale of the processors, like \texttt{fetch\_width} and \texttt{decode\_width}, or describe the microarchitecture decision and technology, for example \texttt{instruction\_window\_scheme} and \texttt{core\_tech\_node}. To collect the event parameters such as \texttt{read\_accesses} of ICache and \texttt{total\_cycles}, 
%Users can adopt performance simulators like gem5~\cite{binkert2011gem5} followed by a parser to transform the statistics of gem5 into XML style. 

% The hardware parameters like \emph{fetch\_width} and \emph{decode\_width} can be set directly by the user. To collect the event parameters such as \emph{read\_accesses} of ICache and \emph{total\_cycles}, users can adopt performance simulators like gem5~\cite{binkert2011gem5} followed by a parser to transform the statistics of gem5 into XML style. For other parameters that further elaborate on the microarchitecture and technology details of the processor, for example \emph{instruction\_window\_scheme} and \emph{core\_tech\_node}, require users to set based on the knowledge of processors. 

With the input XML file, McPAT estimates the power consumption in two steps, as shown in Fig.~\ref{framework}. 
\textbf{Step 1:} With hardware parameters $H$, McPAT instantiates each component based on the CACTI~\cite{li2011cacti} to collect energy consumption of each operation. 
For example, for ICache, McPAT first calculates the configuration, such as cache size \texttt{cache\_sz} and tag width \texttt{tag\_w}, from the hardware parameters \texttt{icache\_config}. Then McPAT uses the configuration to instantiate an \texttt{array} from CACTI. The generated instance provides the energy consumption of an ICache read \texttt{readOp} and an ICache write \texttt{writeOp}. 
\textbf{Step 2:} With event parameters $E$ and energy per operation collected from the first step, McPAT transforms events into operations based on the component microarchitecture and then calculates the component energy. Power is energy divided by execution time. 
Following ICache example, each \texttt{hit} is transformed into a \texttt{readOp}, each \texttt{miss} is transformed into a \texttt{readOp} and a \texttt{writeOp}. Then energy consumption is calculated based on the total number of \texttt{readOp} and \texttt{writeOp} and energy consumption per operation. \looseness=-1

\section{Rethinking: How to Improve Analytical Model}
\label{sec:discrepancy}

%\yao{[Mention previous ML power model and their problem (only calibrate total power). From the methodology perspective]}

% To improve the low accuracy of the traditional analytical power model, ML-based power models are proposed to directly learn the correlation between input features and the final power. 
% However, these ML-based power models that walk around the problems in analytical models incur new problems, including unreliability, limited interpretability, and the lack of a consistent interface.  

%\yao{The real problem of the analytical model.}
% There exist significant discrepancies between the target processor and the original processor assumed in the analytical power model. 
%The design flow of a processor 
%\yao{ .. discrepancies exist in all three major levels, including} 

Observation: The root cause of the low accuracy when applying analytical models to a new processor architecture is \emph{the discrepancies between the real processor implementation and the processor's analytical model}. 
The discrepancies exist in all three major levels, including microarchitecture design, RTL implementation, and technology. 
At each level, different processor architectures may adopt different design options, while the analytical models only support one or very few options, resulting in discrepancies: 
1) For the microarchitecture design, each component has multiple design options. For example, for the ICache, some processors adopt a low-latency design where the tag array and data array are accessed simultaneously to reduce the ICache access latency. Some other low-power processors may access the data array after the tag array access, reducing the power consumption with fewer data array accesses. 
% 2) For the RTL implementation, the same component can be implemented differently, even with similar functions. For example, different implementations may have different widths in the array holding different bits to represent status. As for logic design, it can be even more flexible and lead to different power consumptions~\cite{sengupta2022good}. 
2) For the RTL implementation, the same component can be implemented differently, even with similar functions. For the array structure, continuing with the ICache example, the tag array width may differ in different implementations, because it may hold different numbers of extra bits to represent status. For the logic structure, such as control logic, its design is even more flexible and leads to different power consumption~\cite{sengupta2022good}. 
% 3) For the technology level, the technology node adopted can significantly affect the power consumption, even though the RTL implementation is the same. 
%\yao{[fix the tech discrepency, how analytical model technology]}
3) For the technology, analytical methods model the technology node by setting and scaling empirical built-in values, such as CACTI~\cite{li2011cacti}. 
These empirical values cannot always accurately reflect the real technology library, such as TSMC~\cite{URL:tmsc40nm} across various process nodes.

%Besides, when modeling different technology node such as 28nm and 40nm, the inaccuracy of empirical values will exacerbate.
% Because of the differences across different technology libraries, the analytical model can not always be accurate for every technology library. 
%Different EDA tools can also result in different final designs.

%To improve the low accuracy incurred by the discrepancies, existing ML-based power models~\cite{zhai2022mcpat} adopt a purely data-driven approach to calibrate the analytical models with an ML model. However, the ML-based power model walks around the discrepancies in analytical models by directly learning the correlation between input features and the final power. However, directly fitting the final power with ML models is unreliable incurred to the weak extrapolation. 

Existing ML-based power models~\cite{zhai2022mcpat} are fixing the discrepancy-induced inaccuracies with data-driven approaches. They typically \emph{walk around} the discrepancy problem by directly calibrating the ultimate total power value towards ground-truth with a black-box ML model. As a result, they work well when the test designs fall within the distribution of training designs, but suffer from serious unexpected accuracy degradation when inferring out-of-distribution test data. We demonstrate this with detailed experiments in Section~\ref{subsec:exp-accuracy}.

%instead of fixing discrepancies at each single level. 

%and limits the interpretability because the ML model is a black box. 
%Besides, the from-scratch framework has a significantly different interface from the traditional power model, causing extra difficulty for users.
% directly learn the correlation between input features and the final power. 
% However, these ML-based power models that walk around the problems in analytical models incur new problems, including unreliability, limited interpretability, and the lack of a consistent interface. 

% To tackle the discrepancies above, we regard the analytical power model as a design-flow-aware hierarchy as shown in Fig.~\ref{framework}, including architecture, implementation, and technology level. For each level, we identify important parameters that can be different across different processors and have a significant impact on the power consumption. With these tunable parameters, ReadyPower can bridge the discrepancies between the target processor and the modeled processor.

%as discussed above, we propose ReadyPower.
% Fig.~\ref{framework} shows the framework of ReadyPower.
% ReadyPower regards the analytical power model as a three-level hierarchy as shown in Fig.~\ref{framework}, including architecture, implementation, and technology level.

%between the processor implementation and the model by introducing new parameters into the analytical framework at the architecture, implementation, and technology levels. 

%  for different architectures

Instead of directly calibrating the total power, ReadyPower bridges the discrepancies at each level by introducing new parameters, each with a realistic meaning in power calculation. 
%By capturing different design options at each level, these introduced parameters can bridge the discrepancies between the real target processor architecture and the classical analytical model. 
%In this way, ReadyPower 
These ReadyPower-introduced parameters capture the design option of the real target processor architecture (i.e., BOOM, XiangShan) internally in its analytical framework. Please note that ReadyPower only adopts data-driven methods to decide the values of its introduced new parameters for the target architecture. It is a purely white-box analytical model without an ML component during its power calculation.

%maintaining a high reliability. 

% Compared to the ML-based model that directly fits the final power
% adopted in the processor based on

%\yao{[By capturing different options for different architectures, .....]} % These introduced parameters can bridge the discrepancies between the target processor and the modeled processor. Such an analytical framework can maintain a high reliability, interpretability, and have a consistent interface with existing widely adopted power models.

% For each level, ReadyPower introduces important parameters that can be different across different processors and have a significant impact on the power consumption. With these tunable parameters, ReadyPower can bridge the discrepancies between the target processor and the modeled processor.

\section{Methodology}

\begin{figure}[!t]
\centering
\vspace{-.15in}
\includegraphics[width=0.42\textwidth]{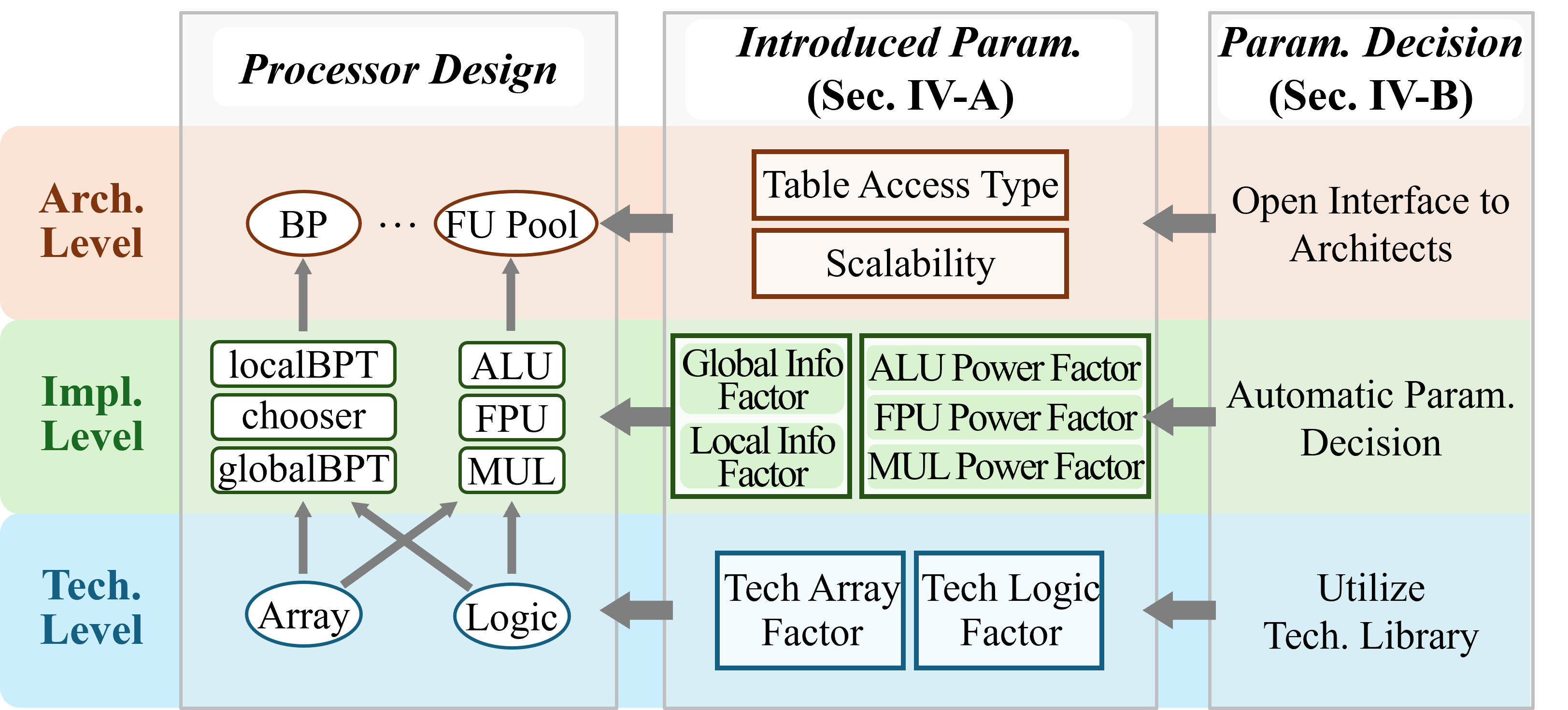}
\vspace{-.05in}
\caption{The methodology overview of ReadyPower.}
\vspace{-.25in}
\label{overview}
\end{figure}

This section introduces the detailed methodology of ReadyPower, with an overview shown in Fig.~\ref{overview}. 
In Sec.~\ref{sec:modelparam}, we elaborate on the introduced parameters at each level. In Sec.~\ref{sec:paramcalib}, we introduce our proposed customized parameter decision methods for each level. We also describe our implementation of ReadyPower in Sec.~\ref{sec:interface}.

% \begin{figure}[!t]
% \centering
% %\vspace{-.1in}
% \includegraphics[width=0.4\textwidth]{_fig/cpuarch.png}
% %\vspace{-.1in}
% \caption{The architecture of our target Out-of-Order CPU core. Blue blocks are key components. The green block indicates the Other Logic.}
% \vspace{-.15in}
% \label{cpuarch}
% \end{figure}

\subsection{Introduced New Parameters}
\label{sec:modelparam}

We introduced architecture-level, implementation-level, and technology-level parameters into the widely adopted McPAT analytical model. In this section, we will first introduce the general out-of-order processor architecture, and then describe introduced new parameters of each level.

\textbf{CPU Architecture: } Before elaborating on our introduced parameters, we first introduce a general CPU core architecture adopted by ReadyPower. % \yao{[simplify]}  %and then elaborate on each level of the parameterization. 
% Fig.~\ref{cpuarch} shows the architecture of our target Out-of-Order CPU core, including 10 major components with blue: Branch Predictor (BP), Instruction Fetch Unit (IFU), Instruction Cache (ICache), Renaming Unit (RNU), Reorder Buffer (ROB), Instruction Schedule Unit (ISU), Register File (Regfile), the pool of functional units (FU Pool), Load Store Unit (LSU), and Data Cache (DCache). Other circuits not covered above are referred to as Other Logic, with green in the figure.
Our target Out-of-Order CPU core includes 10 major components with: Branch Predictor (BP), Instruction Fetch Unit (IFU), Instruction Cache (ICache), Renaming Unit (RNU), Reorder Buffer (ROB), Instruction Schedule Unit (ISU), Register File (Regfile), the pool of functional units (FU Pool), Load Store Unit (LSU), and Data Cache (DCache). Other circuits not covered above are referred to as Other Logic.

\textbf{Architecture-Level Parameter:}
Architecture-level parameters capture major microarchitecture design decisions. %to capture the difference in the microarchitecture design across different processors. 
The introduced architecture-level parameters satisfy two points: 1) The parameters can be easily identified at the microarchitecture design level by architects. 
% The architecture-level parameter should be a part of the interface to architects who model the processor power at the early stage. Therefore, to reduce the burden of use, it should be easily identified by architects and decoupled from the RTL implementation. 
2) The parameters have a significant impact on the power consumption. Therefore, the number of architecture-level parameters can be controlled at an acceptable level for ease of use. 

% \begin{table}[!t]
% \centering
% \resizebox{0.47\textwidth}{!}{
% \begin{tabular}{|c||c|c|}
% \hline
% \multicolumn{1}{|c||}{Component} & \multicolumn{1}{c|}{Architecture-Level Parameter}                                                     & Option                                                      \\ \hline
% \hline
% BP                              & \begin{tabular}[c]{@{}c@{}}Table Access Type\\ Scalability\end{tabular}                              & \begin{tabular}[c]{@{}c@{}}Low Latency / Low Power \\ Yes / No\end{tabular}        \\ \hline
% ICache                          & \begin{tabular}[c]{@{}c@{}}Table Access Type\\ Scalability\end{tabular}                              & \begin{tabular}[c]{@{}c@{}}Low Latency / Low Power \\ Yes / No\end{tabular}        \\ \hline
% DCache                          & Multi-Port Design                                                                                      & Multi-Banking / Duplicated Array                                                       \\ \hline
% \end{tabular}
% }
% \caption{Architecture-Level Parameters.}
% \label{archparam}
% \end{table}

The first part of Table~\ref{table:param} summarizes 5 architecture-level parameters introduced in our framework. % We introduce our introduced architecture-level parameters below. 
1) For the ICache, the processors may adopt a \emph{Low Latency} design or a \emph{Low Power} design for the data array. We refer to its relevant new parameter as \emph{Table Access Type}. 
% In our model implementation, we modify the data array access times to the number of cycles to represent continuous accessing. 
Besides, in some high-performance processors like BOOM~\cite{zhao2020sonicboom}, the ICache line size may scale up with the \emph{FetchWidth} to provide scalable throughput. We refer to it as \emph{Scalability}. 
% In our implementation, we modify the number of ways to represent the throughput scaling up to work around the limited cache model in CACTI~\cite{li2011cacti}. 
2) For the BP, similar to the ICache, as a part of Frontend, it also adopts two architecture-level parameters \emph{Table Access Type} and \emph{Scalability}. 
3) For the DCache, in superscalar processors, it should support multi-port access. The design has multiple options, so we introduce \emph{Multi-Port Design} as a parameter, with two options: \emph{Multi-Banking} and \emph{Duplicated Array}.
% with two options: \emph{Multi-Banking}, the design adopted in XiangShan~\cite{xu2022towards}, and \emph{Duplicated Array}, the design adopted in BOOM~\cite{zhao2020sonicboom}. 
% In our model implementation, we modify the number of banks and the cache size to represent these two designs.

\textbf{Implementation-Level Parameter: }
Implementation-level parameters capture differences across different RTL implementations. The implementation-level parameters mainly tackle three major problems: 
1) For the logic structure, the estimations in analytical models are usually too rough because of the lack of RTL implementation information. Therefore, parameters should be introduced to capture the real logic implementation in the processor. 
2) For the array structure, while it is architecturally visible and well-defined, its actual RTL implementation, including precise width and physical size, remains indeterminate at the architecture level. This uncertainty stems from the multiple implementation options available for any given architectural function.
3) In addition to the two points above, which target parameters related to the hardware design, the inaccuracy of event parameters also introduces the inaccurate prediction of the power model. 
Therefore, we introduce the most important parameters to deal with the event inaccuracy.
% Therefore, customized parameters should be introduced to deal with the event inaccuracy. We only introduce parameters to capture the events that are important for power consumption. 
\looseness=-1

The second part of Table~\ref{table:param} lists our introduced 18 implementation-level parameters. %We introduce our implementation-level parameters below. 
1) For BP, as discussed above, parameters are required to model an equivalent implementation for a variety of processors. Therefore, we set \emph{Global Info Factor} to scale array size of global table and chooser, and \emph{Local Info Factor} to scale array size of local table. 
2) For ICache and DCache, data array size is determined at architecture level, however, the width of tag array should be configurable because processors usually integrate some metadata beyond tag, such as security-related bits and some other bits to represent status. Therefore, we introduce \emph{ICache MetaData Bit} and \emph{DCache MetaData Bit} as the number of additional bits beyond tag. 
3) For IFU, RNU, LSU, and Other Logic, these components have logic as their major circuits. Parameters are required to capture the scale of general logic designs. 
Therefore, we introduce \emph{IFU Logic Factor}, \emph{RNU Logic Factor}, \emph{LSU Logic Factor}, and \emph{Other Logic Factor}. 
4) For Regfile, Regfile in modern processors is physical register file. The width is larger than architectural registers, with some bits to keep information related to, for example, renaming. Therefore, we introduce \emph{Physical Regfile Width} representing the ratio between the real and architectural register width. 
5) For ISU and ROB, status bits are usually included to store instruction status to assist scheduling. Therefore, \emph{Inst. Window Width} and \emph{ROB Entry Width} are introduced to represent status bits. 
6) For FU Pool, the FPU, ALU, and MUL are included for computation. To capture the real power consumption of computation, we introduce \emph{FPU Power Scale}, \emph{ALU Power Scale}, and \emph{MUL Power Scale} to represent the bias between target real design and default design in original analytical model. 
7) For ICache and DCache, to tackle inaccurate event parameters, we scale the number of hits and misses linearly with our introduced \emph{Access Coefficient} and \emph{Access Bias}. \looseness = -1

\begin{table}[!t]
\centering
\vspace{-.05in}
\resizebox{0.46\textwidth}{!}{
\begin{tabular}{|c||c|c|c|c|}
\hline
\multicolumn{1}{|c||}{\textbf{Component}} & \multicolumn{1}{c|}{\textbf{Arch.-Level Param.}}                        & \textbf{Type} & \textbf{Range}                                                                        & \textbf{Default} \\ \hline \hline
ICache                               & \begin{tabular}[c]{@{}c@{}}Table Access Type\\ \\ Scalability\end{tabular}  & \begin{tabular}[c]{@{}c@{}}Enum \\ \\ Bool\end{tabular} & \begin{tabular}[c]{@{}c@{}}Low Latency \\ / Low Power \\ Yes / No\end{tabular}  & \begin{tabular}[c]{@{}c@{}}Low Power \\ \\ No\end{tabular}       \\ \hline
BP                           & \begin{tabular}[c]{@{}c@{}}Table Access Type\\ \\Scalability\end{tabular}  & \begin{tabular}[c]{@{}c@{}}Enum \\ \\Bool\end{tabular} & \begin{tabular}[c]{@{}c@{}}Low Latency \\ / Low Power \\ Yes / No\end{tabular}  & \begin{tabular}[c]{@{}c@{}}Low Power \\ \\ No\end{tabular}       \\ \hline
DCache                           & Multi-Port Design                                                        & Enum & \begin{tabular}[c]{@{}c@{}}Multi-Banking \\ / Duplicated Array\end{tabular}                                             & Multi-Banking       \\ \hline \hline 

\multicolumn{1}{|c||}{\textbf{Component}} & \multicolumn{1}{c|}{\textbf{Impl.-Level Param.}}                                                    & \textbf{Type}                                                          & \textbf{Range}                                                    & \textbf{Default}  \\ \hline
\hline
BP                              & \begin{tabular}[c]{@{}c@{}}Global Info Factor\\ Local Info Factor\end{tabular}                         & \begin{tabular}[c]{@{}c@{}}Int\\ Int\end{tabular}             & \begin{tabular}[c]{@{}c@{}}1-64\\ 1-64\end{tabular}       & \begin{tabular}[c]{@{}c@{}}1\\ 1\end{tabular} \\ \hline
ICache                          & ICache MetaData Bit                                                                                    & Int                                                           & 0-64                                                      & 0 \\ \hline
IFU                             & IFU Logic Factor                                                                                       & Float                                                         & 0-2                                                       & 1 \\ \hline
RNU                             & RNU Logic Factor                                                                                       & Float                                                         & 0-2                                                       & 1 \\ \hline
LSU                             & LSU Logic Factor                                                                                       & Float                                                         & 0-2                                                       & 1 \\ \hline
DCache                          & DCache MetaData Bit                                                                                    & Int                                                           & 0-64                                                      & 0 \\ \hline
Regfile                         & Physical Regfile Width                                                                                 & Int                                                           & 1-16                                                      & 1 \\ \hline
ISU                             & Inst. Window Width                                                                               & Int                                                           & 0-64                                                      & 0 \\ \hline
ROB                             & ROB Entry Width                                                                                        & Int                                                           & 0-64                                                      & 0 \\ \hline
FU Pool                         & \begin{tabular}[c]{@{}c@{}}FPU Power Scale\\ ALU Power Scale\\ MUL Power Scale\end{tabular}            & \begin{tabular}[c]{@{}c@{}}Float\\ Float\\ Float\end{tabular} & \begin{tabular}[c]{@{}c@{}}0-16\\ 0-16\\ 0-16\end{tabular} & \begin{tabular}[c]{@{}c@{}}1\\ 1\\ 1\end{tabular} \\ \hline
Other Logic                     & Other Logic Factor                                                                                     & Float                                                         & 0-2                                                       & 1 \\ \hline \hline
ICache                          & \begin{tabular}[c]{@{}c@{}}Access Coefficient\\ Access Bias\end{tabular}                               & \begin{tabular}[c]{@{}c@{}}Float\\ Float\end{tabular}         & \begin{tabular}[c]{@{}c@{}}0-32\\ 0-32\end{tabular}       & \begin{tabular}[c]{@{}c@{}}1\\ 0\end{tabular} \\ \hline
DCache                          & \begin{tabular}[c]{@{}c@{}}Access Coefficient\\ Access Bias\end{tabular}                               & \begin{tabular}[c]{@{}c@{}}Float\\ Float\end{tabular}         & \begin{tabular}[c]{@{}c@{}}0-32\\ 0-32\end{tabular}       & \begin{tabular}[c]{@{}c@{}}1\\ 0\end{tabular} \\ \hline \hline

\multicolumn{1}{|c||}{\textbf{Circuit}} & \multicolumn{1}{c|}{\textbf{Tech.-Level Param.}}   & \textbf{Type}   & \textbf{Range}   & \textbf{Default}  \\ \hline \hline
Logic                          & Tech Logic Factor                                 & Float  & -  & 1        \\ \hline
Array                          & Tech Array Factor                                 & Float  & -  & 1        \\ \hline
\end{tabular}
}
\vspace{-.05in}
\caption{Our introduced new parameters in ReadyPower, including 5 architecture-level parameters, 18 implementation-level parameters, and 2 technology-level parameters.}
\vspace{-.2in}
\label{table:param}
\end{table}

\textbf{Technology-Level Parameter:}  
Technology-level parameters represent the difference among different technology libraries. Although analytical models such as McPAT~\cite{li2009mcpat} account for technology library differences, their estimations rely on empirical knowledge and thus lack sufficient accuracy. As a result, detailed technology-level parameters are still necessary.

However, capturing all details of the technology library is infeasible. We observe that logic cells generally follow one consistent scaling trend, while array macros adhere to another. To incorporate the impact of the technology node while maintaining model simplicity, we represent its effect using 2 parameters, as shown in the third part of Table~\ref{table:param}: \emph{Tech Logic Factor} and \emph{Tech Array Factor}. 1) \emph{Tech Logic Factor} represents the ratio of logic power in the target real technology node to that of the originally modeled node in the analytical model. 2) \emph{Tech Array Factor} represents the analogous ratio for array power. Note that for all processors implemented in the same technology library, these factors remain identical. 
% representing the ratio of logic power in the target real technology node to that of the originally modeled node in the analytical model, and \emph{Tech Array Factor} representing the analogous ratio for array power. Note that for all processors implemented in the same technology library, these factors remain identical. 

\subsection{Parameter Decision}
\label{sec:paramcalib}

%to achieve both high accuracy and ease of use. 

Building on our proposed analytical power model with our introduced new parameters, ReadyPower provides customized parameter decision methods for each level. 
1) For architecture-level parameters that directly reflect microarchitectural decisions, we open the interface for architects' input. 2) For implementation-level parameters, similar to ML-based methods, parameter values are decided automatically by training with a few known design configurations. 3) For technology-level parameters, parameters are decided automatically based on the technology library data. Parameters are decided in the order of: Architecture $\rightarrow$ Technology $\rightarrow$ Implementation.

\textbf{Architecture-Level Decision: }
Architecture-level parameters are visible to architects. Therefore, we open the interface to architects, allowing them to configure architecture-level parameters regarding their target processors. The interface is an extension to the original hardware parameters in the McPAT input XML file. As discussed in Sec.~\ref{sec:modelparam}, we only identify the major microarchitecture decisions that have a significant impact on power consumption, so the number of parameters is acceptable for architects to configure, maintaining the ease of use while ensuring the correctness of these parameters.

\textbf{Implementation-Level Decision:} For the implementation-level parameters, we propose an automatic parameter decision algorithm, based on a few known configurations of the target processor architecture as training data. We decide the implementation-level parameters for each component independently.

\emph{Problem Formulation:} Given known configuration-workload combinations as training data, we have their hardware parameters $H$ and event parameters $E$ for each component, as power model inputs. We then collect the ground-truth power label $L$ of each component. %Each combination can be represented as $((H, E), L)$. 
We denote the number of implementation-level parameters as $N$, the $i$-th parameter as $p_i$, and the analytical power model as $f$ with hardware parameters $H$ and event parameters $E$ as input. 
The goal of this automatic parameter decision algorithm is to find the implementation-level parameters $\{p_1, ..., p_i, ..., p_N\}$ that can minimize the error between the model prediction $P$ denoted as Eq.(\ref{eq:pred}) and the ground truth power $L$. 
%To minimize the error, we minimize the loss function $Loss$ denoted as Eq.(\ref{eq:loss}).
We adopt the MSE error as our loss function, as shown in Eq.(\ref{eq:loss}). \looseness=-1
\begin{equation}
P = \,f(H, E, p_1, ..., p_i, ..., p_N) \label{eq:pred}
\end{equation}
\begin{equation}
\text{Loss} = \,(P - L)^2 \label{eq:loss} 
\end{equation}

% 我们插入了一些参数，最后我们发现...
% 中间结果和最终结果的关系

% \emph{Observation: }
% We find that, in most analytical models, the power consumption scales linearly and differentiably with the implementation-level parameters coupled with logic-dominated components, while non-linearly and non-differentiably with parameters of array-dominated components. (1) For the logic-dominated components, in most analytical power models, it usually estimates the number of pipeline registers, then multiplies it by an empirical factor, for example, 1.5 in McPAT, to estimate the scale of total logic, and finally adopts the high-level technology library to estimate the power consumption, where the calculation is simple and usually adopts a linear correlation to the logic scale. The parameters coupled with logic-dominated components are additional float factors, which will be multiplied by the original empirical factor to accurately estimate the logic scale. Therefore, such a simple calculation leads to a linear and differentiable correlation between parameters and the power estimation. (2) For the array-dominated components, the integer parameters usually determine the shape of the arrays. Then the shape will be fed into the high-level library, such as CACTI in McPAT, to estimate power information. However, the calculation in the library is usually complex. For example, it breaks the shape into multiple small SRAM cells and then sums them up. Therefore, the correlation between parameters and the power estimation is usually non-linear and non-differentiable. 

\begin{table*}[!t]
\centering
\vspace{-.15in}
      %\vspace{-.35in}
    %  \renewcommand{\arraystretch}{1.05}
      \resizebox{1\textwidth}{!}{
        \begin{tabular}{ |c||c c c c c c c c c c c c c c c||c c c c c c c c c c| } 
\hline
Hardware Parameter  & B1 & B2 & B3 & B4 & B5 & B6 & B7 & B8 & B9 & B10 & B11 & B12 & B13 & B14 & B15 & X1 & X2 & X3 & X4 & X5 & X6 & X7 & X8 & X9 & X10\\
\hline
\hline
FetchWidth & 4 & 4 & 4 & 4 & 4 & 8 & 8 & 8 & 8 & 8 & 8 & 8 & 8 & 8 & 8 &      4 & 4 & 4 & 4 & 4 & 8 & 8 & 8 & 8 & 8\\
\hline
DecodeWidth & 1 & 1 & 1 & 2 & 2 & 2 & 3 & 3 & 3 & 4 & 4 & 4 & 5 & 5 & 5 &     2 & 2 & 2 & 3 & 3 & 3 & 4 & 4 & 4 & 5\\
\hline
FetchBufferEntry & 5 & 8 & 16 & 8 & 16 & 24 & 18 & 24 & 30 & 24 & 32 & 40 & 30 & 35 & 40 &    8 & 16 & 24 & 16 & 24 & 24 & 24 & 32 & 32 & 24\\
\hline
RobEntry & 16 & 32 & 48 & 64 & 64 & 80 & 81 & 96 & 114 & 112 & 128 & 136 & 125 & 130 & 140 &     16 & 32 & 48 & 64 & 64 & 80 & 81 & 96 & 114 & 112\\
\hline
IntPhyRegister & 36 & 53 & 68 & 64 & 80 & 88 & 88 & 110 & 112 & 108 & 128 & 136 & 108 & 128 & 140 &     36 & 53 & 68 & 64 & 80 & 88 & 88 & 110 & 112 & 108\\
\hline
FpPhyRegister & 36 & 48 & 56 & 56 & 64 & 72 & 88 & 96 & 112 & 108 & 128 & 136 & 108 & 128 & 140 &     36 & 53 & 68 & 64 & 80 & 88 & 88 & 110 & 112 & 108\\
\hline
LDQ/STQEntry & 4 & 8 & 16 & 12 & 16 & 20 & 16 & 24 & 32 & 24 & 32 & 36 & 24 & 32 & 36 &     16 & 20 & 24 & 20 & 24 & 28 & 24 & 32 & 40 & 32\\
\hline
BranchCount & 6 & 8 & 10 & 10 & 12 & 14 & 14 & 16 & 16 & 18 & 20 & 20 & 18 & 20 & 20 &     7 & 7 & 7 & 7 & 7 & 7 & 7 & 7 & 7 & 7\\
\hline
Mem/FpIssueWidth & 1 & 1 & 1 & 1 & 1 & 1 & 1 & 1 & 2 & 1 & 2 & 2 & 2 & 2 & 2 &     2 & 2 & 2 & 2 & 2 & 2 & 2 & 2 & 2 & 2\\
\hline
IntIssueWidth & 1 & 1 & 1 & 1 & 2 & 2 & 2 & 3 & 3 & 4 & 4 & 4 & 5 & 5 & 5 &     2 & 2 & 2 & 2 & 4 & 4 & 4 & 6 & 6 & 6\\
\hline
DCache/ICacheWay & 2 & 4 & 8 & 4 & 4 & 8 & 8 & 8 & 8 & 8 & 8 & 8 & 8 & 8 & 8 &     4 & 4 & 8 & 4 & 4 & 8 & 8 & 8 & 8 & 8\\
\hline
DTLBEntry & 8 & 8 & 16 & 8 & 8 & 16 & 16 & 16 & 32 & 32 & 32 & 32 & 32 & 32 & 32 &     8 & 8 & 16 & 8 & 8 & 16 & 16 & 16 & 32 & 32\\
\hline
MSHREntry & 2 & 2 & 4 & 2 & 2 & 4 & 4 & 4 & 4 & 4 & 4 & 8 & 8 & 8 & 8 &     2 & 2 & 4 & 2 & 2 & 4 & 4 & 4 & 4 & 4\\
\hline
ICacheFetchBytes & 2 & 2 & 2 & 2 & 2 & 4 & 4 & 4 & 4 & 4 & 4 & 4 & 4 & 4 & 4 &     2 & 2 & 2 & 2 & 2 & 2 & 2 & 2 & 2 & 2\\
         \hline
        \end{tabular}
        }
       \vspace{-.03in}
        %\vspace{-1mm}
        \caption{The CPU configurations used in our experiment. The B1-B15 denote the 15 configurations of BOOM, and the X1-X10 denote the 10 configurations of XiangShan. The scales of these configurations are from small to large.}
        \vspace{-.17in}
        \label{configtable}
\end{table*}

\emph{Automatic Parameter Decision Algorithm: }
We derive an automatic parameter decision algorithm based on gradient descent, as shown in Eq.(\ref{eq:update})(\ref{eq:gradient})(\ref{eq:arraygrad}). For each iterative step during training, parameter $p_i$ is updated based on its partial gradient and learning rate $lr$, as shown in Eq.(\ref{eq:update}). 
%\footnote{When updating the integer $p_i$, we round up positive $lr*\frac{\partial L}{\partial p_i}$ and round down negative ones.}. 
Partial gradient with respect to $p_i$ is calculated in Eq.(\ref{eq:gradient}). 
% \footnote{We use the widely adopted learning rate decay in our algorithm to adjust $lr$. To be specific, the learning rate of the $k$-th iteration $lr_k$ can be expressed as $lr_k = lr_0 * d^{k}$, where $d\,(d<1)$ is the decay rate.}. % \looseness=-1  
\begin{equation}
\forall i \in [i, N], \ \  p_i \leftarrow \, p_i - lr * \frac{\partial \text{Loss}}{\partial p_i} \label{eq:update}
\end{equation}
\begin{equation}
\frac{\partial \text{Loss}}{\partial p_i} = \,2(P - L)\frac{\partial P}{\partial p_i} \label{eq:gradient}
\end{equation}
\noindent The task is to calculate the $\frac{\partial P}{\partial p_i}$ in Eq.(\ref{eq:gradient}). As indicated in Eq.(\ref{eq:pred}), power $P$ is calculated based on the analytical power model $f$, which is not directly differentiable. To solve this, we approximate the $\frac{\partial P}{\partial p_i}$ by slightly changing the parameter $p_i$ by a tiny $\delta$ and calculating the slope. Specifically, we first run the analytical model $f$ to collect the prediction, denoted as $f(H, E, p_1, ..., \boldsymbol{p_i}, ..., p_N)$, and then increase $p_i$ by a tiny $\delta$ to collect the new prediction, denoted as $f(H, E, p1, ..., \boldsymbol{p_i + \delta}, ..., p_N)$, finally we approximate $\frac{\partial P}{\partial p_i}$ as below:  \looseness=-1
\begin{equation}
\frac{\partial P}{\partial p_i} = \frac{f (H, E, p1, ..., \boldsymbol{p_i + \delta}, ..., p_N) - f(H, E, p_1, ..., \boldsymbol{p_i}, ..., p_N)}{\delta}   \label{eq:arraygrad} 
\end{equation}
Combining Eq.(\ref{eq:update})(\ref{eq:gradient})(\ref{eq:arraygrad}), we can decide all parameters with gradient descent, through multiple iterations. 
In addition, some introduced parameters $p_i$ are actually linearly correlated with the analytical model's power estimation $P$. 
For these linear parameters, the $\frac{\partial P}{\partial p_i}$ is unchanged across iterations and only needs calculation once.

\iffalse

\begin{align}
& p_i \leftarrow \, p_i - lr * \frac{\partial \text{Loss}}{\partial p_i} \label{eq:update} \\ 
%
& \frac{\partial \text{Loss}}{\partial p_i} = \,2(P - L)\frac{\partial P}{\partial p_i} \label{eq:gradient} \\
%
& \frac{\partial P}{\partial p_i} = \frac{P^{p_i + \delta} - P^{p_i}}{\delta}   \label{eq:arraygrad} 
\end{align}

\fi

\textbf{Technology-Level Decision: }
To decide the technology-level parameters, including \emph{Tech Logic Factor} and \emph{Tech Array Factor}, we utilize the information from the technology library. Only these two parameters are introduced to maintain model simplicity. 

%To maintain model simplicity, we capture the effect of technology using only two parameters.

%for an accurate decision.

\emph{Tech Array Factor:} To determine the \emph{Tech Array Factor}, we generate a basic SRAM macro with memory compiler in the technology library, for example, a single-port SRAM with shape 256$\times$64, and collect its read and write energy. Then, we create the same array with CACTI in the McPAT, and collect its read and write energy. Finally, we can calculate the ratio between the read energy of the SRAM macro generated by the memory compiler and the read energy of the array from the analytical model, and the ratio for write energy. We take the average between these two ratios as our \emph{Tech Array Factor}. 

\emph{Tech Logic Factor:} Analytical models usually estimate the logic power based on the power estimation of the register, including clock pin power, switch power, and power consumption for keeping 0 and 1. For example, in McPAT, the worst-case power of DFF, where the clock is toggling, is estimated as a base unit for pipeline logic power estimation. Therefore, to align the worst-case power of DFF, we extract the real worst-case power of DFF from the technology library and the original estimation of the worst-case power from McPAT. The ratio between the two values is our decided \emph{Tech Logic Factor}.

% \qijun{(Model Implementation and User Interface)}

\subsection{Implementation and Interface}
\label{sec:interface}

Finally, we summarize the implementation of our open-source ReadyPower based on the analytical framework. To introduce our new parameters into the McPAT, we directly modify the source code of McPAT to inject these parameters into the calculation flow of the original McPAT. 
1) Architecture-level parameters are injected by modifying the transformation from event to operation. Some others are injected by instrumenting hardware parameters. 
2) Implementation-level parameters are encoded by modifying the configuration provided for the circuit instantiation. Event-related ones are injected by scaling event parameters before transformation into operations. 
% When performing the automatic parameter decision algorithm, for each iteration, our ReadyPower framework will automatically modify the McPAT code with updated parameters and recompile it. Each compilation takes only a few seconds. 
3) We inject technology-level parameters by scaling the operation energy when instantiating circuits with CACTI.

With such a source-code level parameter injection, ReadyPower can maintain the identical user interface (with several additional architecture-level parameters) to the original McPAT. %, taking the XML file with hardware and event parameters as input and outputting the power estimation.

\section{Experimental Results}

\subsection{Experiment Setup}

In our experiment, we adopt two mainstream open-sourced Out-of-order CPU architectures, BOOM~\cite{zhao2020sonicboom} and XiangShan~\cite{xu2022towards}, which have been widely adopted in prior works~\cite{zhai2022mcpat,zhai2023microarchitecture,zhang2023panda}, to evaluate our proposed solution. We generate 15 different configurations of BOOM CPU (B1-B15) and 10 configurations of XiangShan CPU (X1-X10) in our experiments, as shown in Table~\ref{configtable}. To simulate real workloads, we adopt eight workloads in riscv-tests~\cite{URL:riscvtests}, including dhrystone, median, multiply, qsort, resort, towers, spmv, and vvadd. 

%Minor code modifications are made to execute on the XiangShan CPU. \looseness=-1

To collect the ground truth power labels, we perform RTL code generation and RTL simulation of BOOM with Chipyard framework~\cite{amid2020chipyard} v1.8.1 and XiangShan with the OpenXiangShan framework~\cite{xu2022towards}, where Synopsys VCS\textsuperscript{\textregistered}~\cite{vcs} is utilized. We then perform logic synthesis with Synopsis Design Compiler\textsuperscript{\textregistered}~\cite{design-compilier}. We turn on the clock-gating option, which is often missed in prior works, during logic synthesis. The post-synthesis power simulation is performed with PrimePower~\cite{ptpx}. The whole synthesis and simulation flow is performed based on TSMC 40nm standard cell library~\cite{URL:tmsc40nm} and the associated Memory Compiler for SRAM generation. TSMC 28nm library~\cite{URL:tmsc28nm} is additionally adopted for the cross-technology-node transfer experiment. To collect the event parameters, we perform architecture-level performance simulation based on gem5~\cite{binkert2011gem5}.

% Our base analytical power model is McPAT~\cite{li2009mcpat}.

\subsection{Training and Testing Data Setup}
We introduce our setup of training and testing data in this section. 
To reflect the real scenarios where limited configurations are available, we set the number of available training configurations to 3. We evaluate each method under three training scenarios with different training data distributions: 
1) \emph{Balance}. We evenly select the configurations as available training configurations based on the scale: B1, B8, and B15 for BOOM, X1, X6, and X10 for XiangShan. 
2) \emph{Small}. We select the smallest configurations as available training configurations: B1, B2, and B3 for BOOM, X1, X2, and X3 for XiangShan. 
3) \emph{Large}. We select the largest configurations as available training configurations: B13, B14, and B15 for BOOM, X8, X9, and X10 for XiangShan. 
For each training scenario, all remaining configurations are for testing. 
In \emph{Small} and \emph{Large} training scenarios, the distribution of testing data falls out of the distribution of training data: 
For \emph{Small} training scenario, \emph{all} testing configurations are \emph{larger} than training configurations. For \emph{Large} training scenario, \emph{all} testing configurations are \emph{smaller} than training configurations. 
Because of the difference between the distribution of training and testing data, these two scenarios are very challenging and usually ignored by prior works~\cite{zhai2022mcpat,zhai2023microarchitecture}. 
However, these two training scenarios are important and realistic because architects usually also work on configurations that have different scales from available known configurations.
%However, these two training scenarios are important and realistic because architects prefer to elaborate on configurations with a similar scale in the initial design stage, therefore, only small configurations or only large configurations are available. 
% 强调非常challenging，在result里面也强调
% 所有的都比training大或者小
% 实际上会做与已有design不同规模的config

\subsection{Summary of Baseline Methods}
We compare ReadyPower with representative architecture-level power models as our baseline, including (a) the analytical power model McPAT~\cite{li2009mcpat} and (b) the ML-based power model McPAT-Calib~\cite{zhai2022mcpat}. Besides the two baselines proposed in prior works, we also include two extra baselines: (c) McPAT-Plus, an enhanced analytical model. We derive a scaling factor based on the prediction and ground-truth of the available training configurations, and scale the power prediction of McPAT for testing. (d) McPAT-Calib-Component, an enhanced ML-based model. It builds ML-based models for each component, with related hardware parameters and event parameters as features and per-component power as the label. XGBoost~\cite{chen2016xgboost}, the best algorithm reported by McPAT-Calib~\cite{zhai2022mcpat}, is adopted as the algorithm for ML-based models.

ASP-DAC'23~\cite{zhai2023microarchitecture} and PANDA~\cite{zhang2023panda} are beyond the scope of this work. 
ASP-DAC'23~\cite{zhai2023microarchitecture} is a transfer learning method, explicitly requiring training and testing configurations to be similar to enable transfer learning.
PANDA~\cite{zhang2023panda} requires significant engineering effort to analyze the design and develop resource functions for each component, which is not an automated power modeling method. 

% We exclude ASP-DAC'23~\cite{zhai2023microarchitecture} and PANDA~\cite{zhang2023panda} for comparison: 
% ASP-DAC'23~\cite{zhai2023microarchitecture} requires available configurations across all scales. 
% PANDA~\cite{zhang2023panda} requires significant engineering effort to analyze the design and develop resource functions for each component. 
% 提一下PANDA, FirePower, ASPDAC'23

% \subsection{Evaluation Scenarios}
% To reflect the real scenarios where limited configurations are available, we set the number of available training configurations to 3. We evaluate each method under three scenarios with different training data distributions. (1) Balance. We evenly select the configurations as available training configurations based on the scale: B1, B8, and B15 for BOOM, X1, X6, and X10 for XiangShan. (2) Small. We select the smallest configurations as available training configurations: B1, B2, and B3 for BOOM, X1, X2, and X3 for XiangShan. (3) Large. We select the largest configurations as available training configurations: B13, B14, and B15 for BOOM, X8, X9, and X10 for XiangShan. 

\subsection{Power Modeling Accuracy}
\label{subsec:exp-accuracy}

\begin{figure}[!t]
\centering
\vspace{-.05in}
\includegraphics[width=0.42\textwidth]{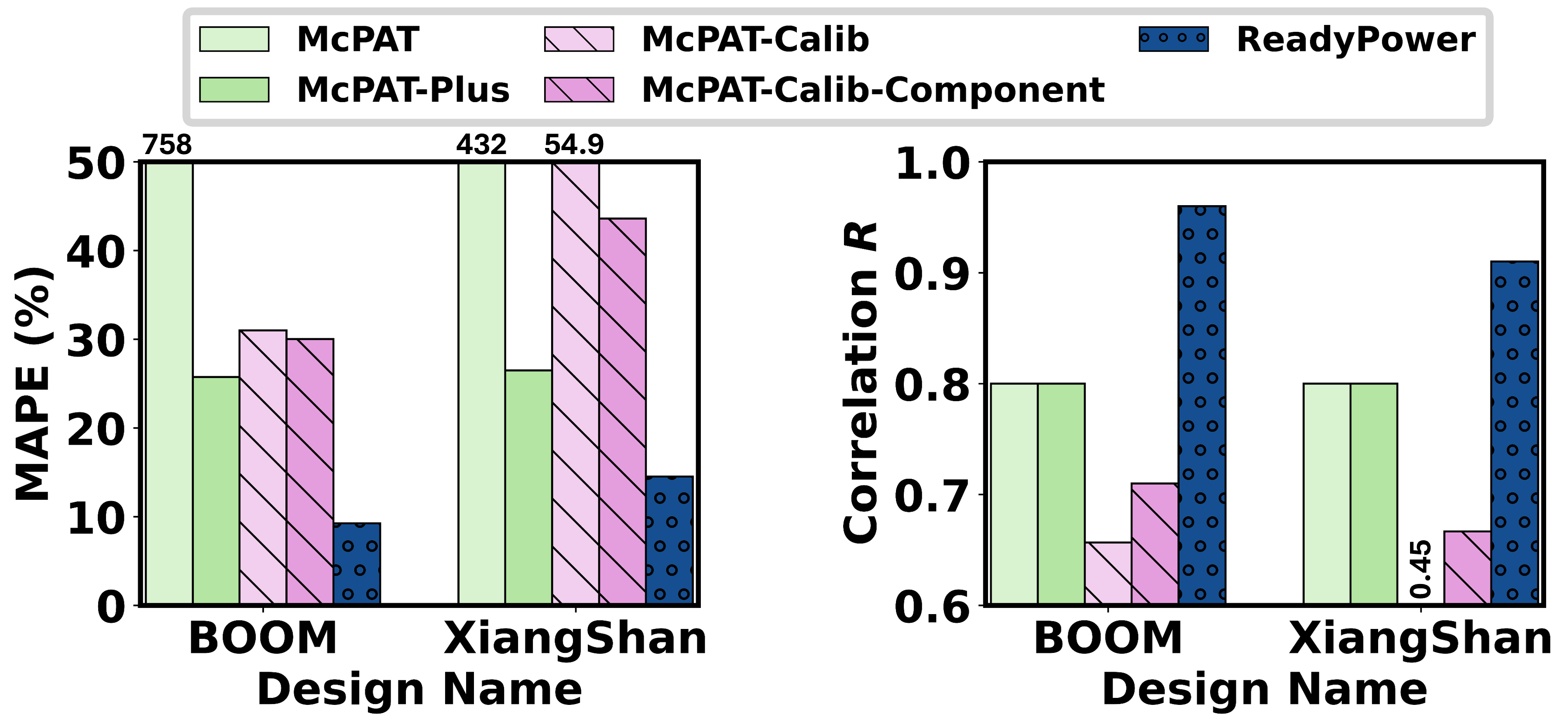}
\vspace{-.05in}
\caption{Accuracy comparison between our proposed ReadyPower and four baseline methods on BOOM and XiangShan CPUs. Each bar is the average across all training scenarios, including \emph{Balance}, \emph{Small}, and \emph{Large}.\looseness=-1}
\vspace{-.1in}
\label{avg}
\end{figure}

Fig.~\ref{avg} demonstrates the accuracy comparison between ReadyPower and our four baselines. The data shown is the average value across \emph{Balance}, \emph{Small}, and \emph{Large} training scenarios. It shows that our proposed ReadyPower can significantly outperform all baseline methods, including existing analytical power models and ML-based power models, on both BOOM and XiangShan. Averaged across all training scenarios, ReadyPower can achieve the lowest MAPE (mean absolute percentage error) of 9.26\% and 14.51\%, and the highest correlation coefficient $R$ of 0.96 and 0.91 on BOOM and XiangShan. %\looseness=-1

\begin{figure}[!t]
\centering
\vspace{-.1in}
\hspace{-2mm}
\includegraphics[width=0.5\textwidth]{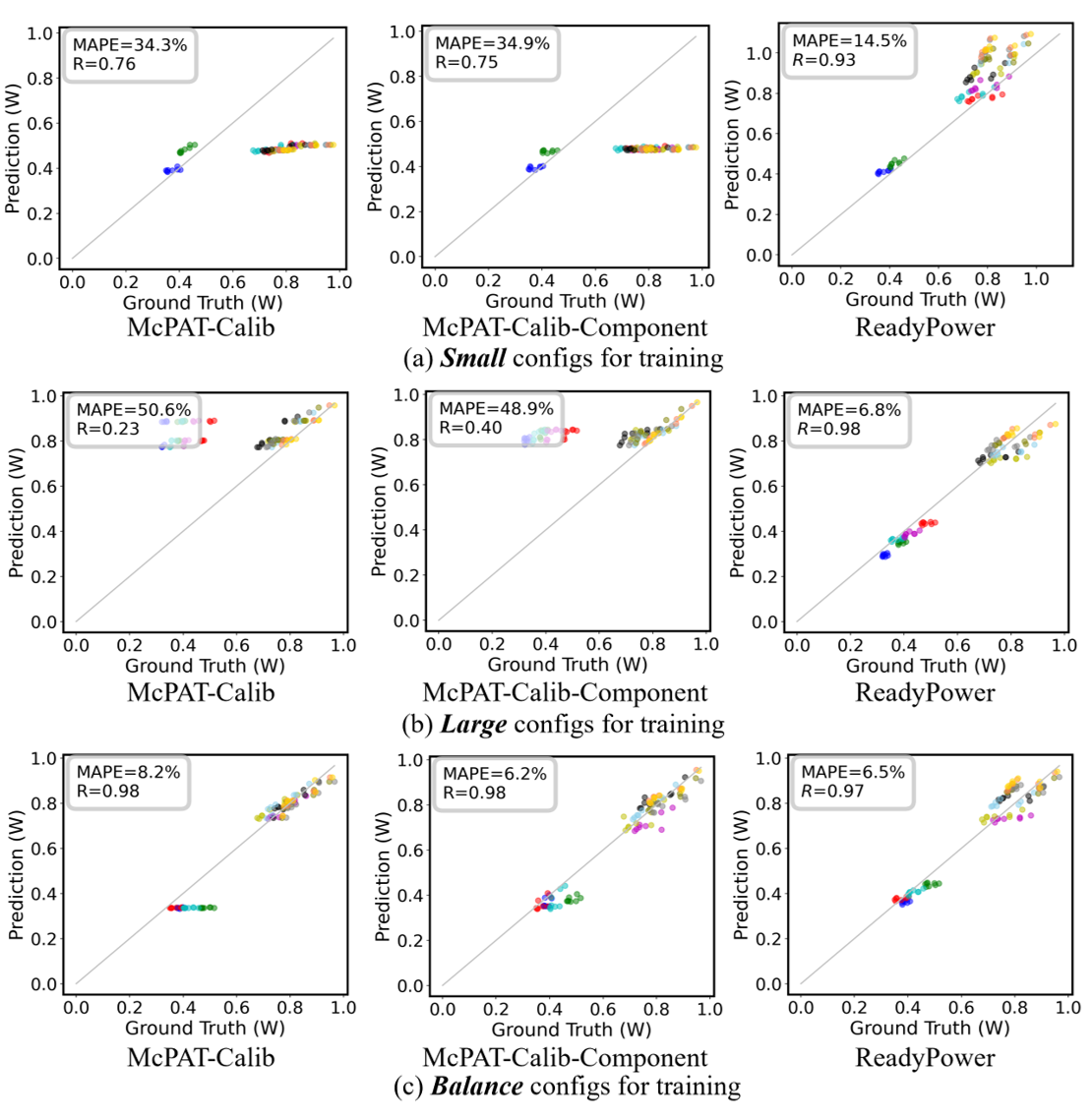}
\vspace{-.25in}
\caption{Prediction visualization on BOOM CPU. It shows that ReadyPower can achieve high accuracy in ALL training scenarios while ML-based methods are inaccurate in \emph{Small} and \emph{Large} training scenarios on BOOM.}
\vspace{-.2in}
\label{boomexp}
\end{figure}

\begin{figure}[!t]
\centering
\vspace{-.05in}
\hspace{-2mm}
\includegraphics[width=0.5\textwidth]{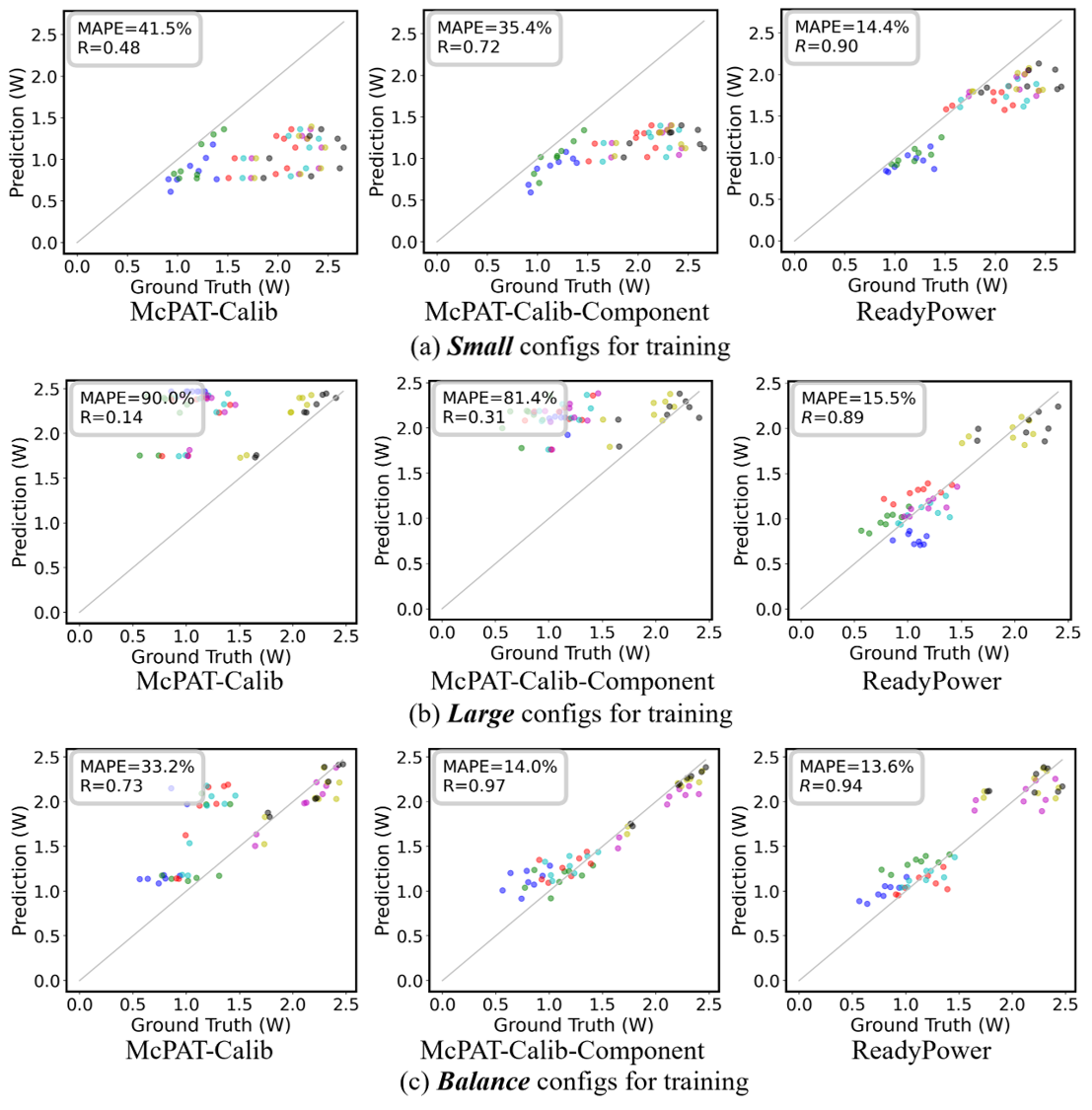}
\vspace{-.25in}
\caption{Prediction visualization on XiangShan CPU. It shows ReadyPower can achieve high accuracy in ALL training scenarios while ML-based methods are inaccurate in \emph{Small} and \emph{Large} scenarios on XiangShan.}
\vspace{-.2in}
\label{xsexp}
\end{figure}

In detail, as shown in the figure, compared with the representative analytical method, the original McPAT and the enhanced version, McPAT-Plus, ReadyPower can consistently achieve much higher accuracy on both architectures, with a 16.49\% and 11.99\% lower MAPE and 0.16 and 0.11 higher correlation $R$. It demonstrates that the introduced new parameters and decisions in ReadyPower can improve the accuracy of the conventional analytical model, McPAT. 
Compared with the representative ML-based method, ReadyPower can surpass the McPAT-Calib and even McPAT-Calib-Component, with a 20.76\% and 29.10\% lower MAPE, and 0.25 and 0.24 higher correlation $R$ on the two architectures. 
% ML-based methods have bad accuracy in \emph{Small} and \emph{Large} training cases, so the average accuracy across these three training cases is very low. 
Such a superior accuracy over ML-based methods shows that ReadyPower is more reliable than the ML-based models under different training data distributions.

Fig.~\ref{boomexp} and Fig.~\ref{xsexp} provide more detailed prediction comparisons between ReadyPower and our baseline methods on BOOM and XiangShan CPUs, respectively. The detailed comparisons further demonstrate the unreliability of ML-based methods and the reliability of ReadyPower. 
% Fig.~\ref{boomexp} and Fig.~\ref{xsexp} visualize the comparison between ReadyPower and our baseline methods on BOOM and XiangShan, respectively. 
Each point in the figure represents a configuration running a workload. Points with the same color are workloads on the same configuration. The gray line means the accurate prediction.

% while the predictions in the \emph{Balance} training scenario have high accuracy,

As shown in Fig.~\ref{boomexp}(a)(b) and Fig.~\ref{xsexp}(a)(b), ML-based methods, both McPAT-Calib and McPAT-Calib-Component, are significantly inaccurate in \emph{Small} and \emph{Large} training scenarios, in comparison with their accurate predictions in the \emph{Balance} training scenario.  %are significantly inaccurate.  
In the (a) \emph{Small} training scenario, as we expected, ML-based methods tend to underestimate the power consumption on testing data. This is because the model is only exposed to designs with lower power during training. In the (b) \emph{Large} training scenario, for similar reasons, ML-based methods exhibit overestimation on testing data. 
% ML-based methods tend to underestimate the power consumption on testing data in the \emph{Small} training scenario and overestimate the power consumption in the \emph{Large} training scenario. 
This validates the unreliability of ML-based methods, which rely on the similarity between the distribution of training and testing data. 
Because of the low accuracy in \emph{Small} and \emph{Large} training scenarios, the average accuracy of ML-based methods across these three training scenarios, shown in Fig.~\ref{avg}, is very low.
% This is because ML-based methods significantly rely on the similarity between the distribution of training and testing data, leading to unreliable predictions: ML-based methods tend to underestimate the power consumption on testing data in the \emph{Small} training case and overestimate the power consumption in the \emph{Large} training case. 
In comparison, ReadyPower can achieve high accuracy for all three training scenarios with different training data distributions, 
including not only \emph{Balance} but also challenging \emph{Small} and \emph{Large}, 
% including \emph{Small}, \emph{Large}, and \emph{Balance}, 
as shown in Fig.~\ref{boomexp}(a)(b)(c) and Fig.~\ref{xsexp}(a)(b)(c). This demonstrates the reliability of ReadyPower, which can consistently make accurate predictions and is not affected by the distribution gap between training and testing data.
\looseness=-1

As for the runtime, ReadyPower is very fast for both inference and training. The inference time of ReadyPower is within 15 seconds on our server, which is a similar speed to the original McPAT. The training time is within 10 minutes. Notice that only one training process is required for each architecture, and then it can be used for different configurations and different workloads. 

\begin{figure}[!t]
\centering
\vspace{-.1in}
\includegraphics[width=0.42\textwidth]{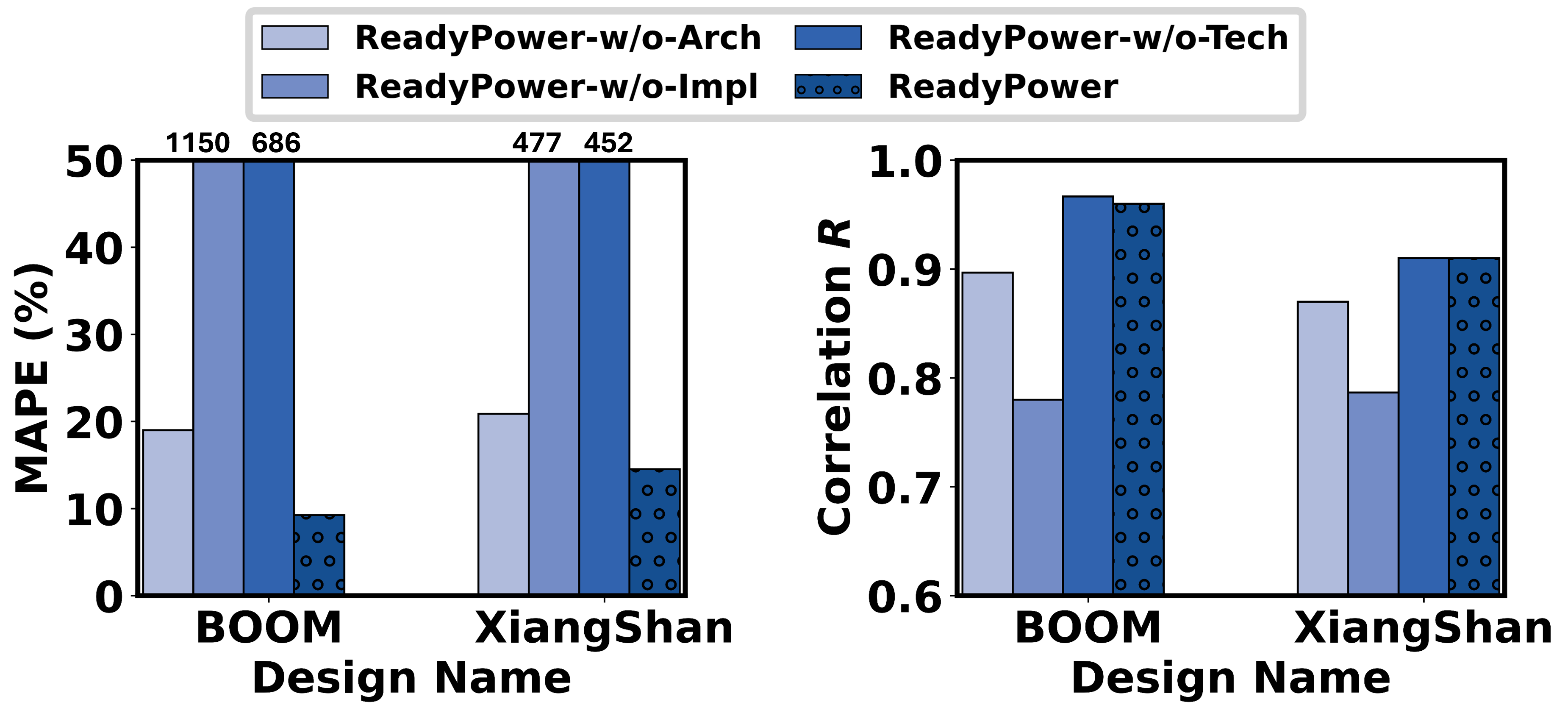}
\vspace{-.05in}
\caption{Ablation studies of ReadyPower. We derive ablation studies, including ReadyPower-w/o-Arch, ReadyPower-w/o-Impl, and ReadyPower-w/o-Tech, for each level in ReadyPower, setting architecture, implementation, and technology-level parameters to default, respectively. Each bar is the average value across all training scenarios. }
\vspace{-.2in}
\label{ablation}
\end{figure}

\begin{figure}[!t]
\centering
\vspace{-.05in}
\hspace{-5mm}
\subfigure[ReadyPower-w/o-Arch]{
    \centering
    \includegraphics[height=0.165\textwidth]{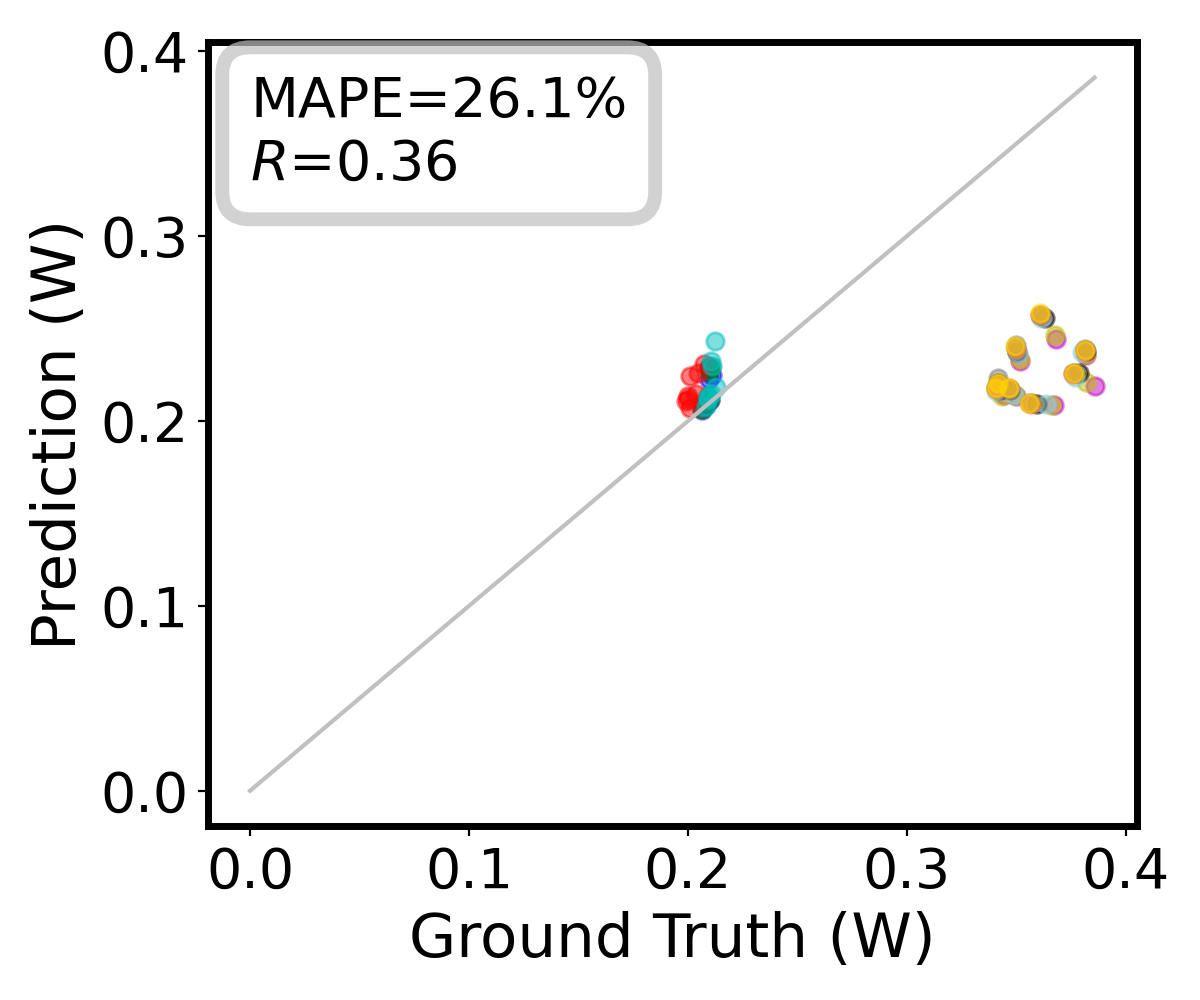}
    \label{archbp}
}
\hspace{-2mm}
\subfigure[ReadyPower-w/o-Impl]{
    \centering
    \includegraphics[height=0.165\textwidth]{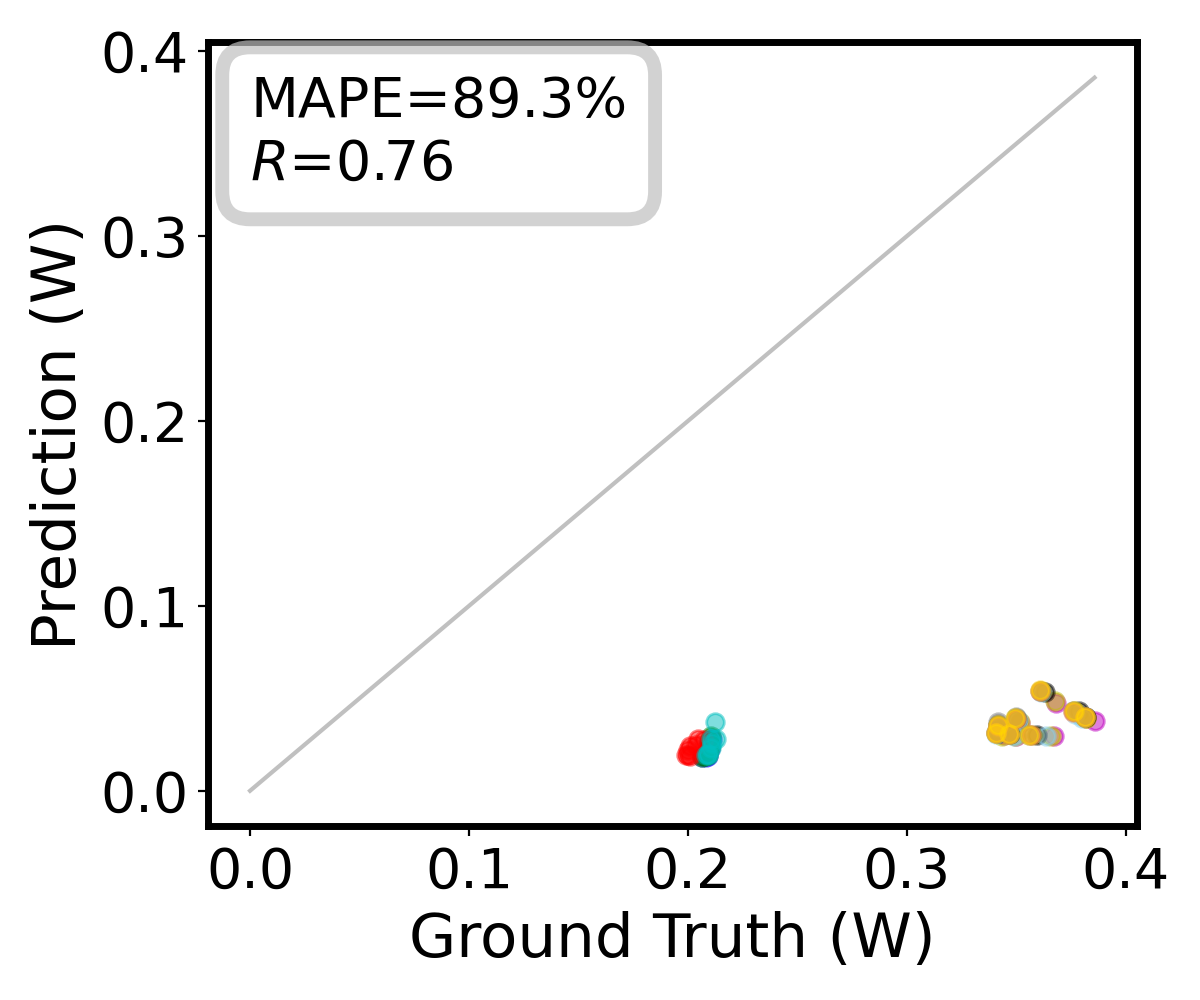}
    \label{implbp}
}

\vspace{-.1in}
\hspace{-5mm}
\subfigure[ReadyPower-w/o-Tech]{
    \centering
    \includegraphics[height=0.165\textwidth]{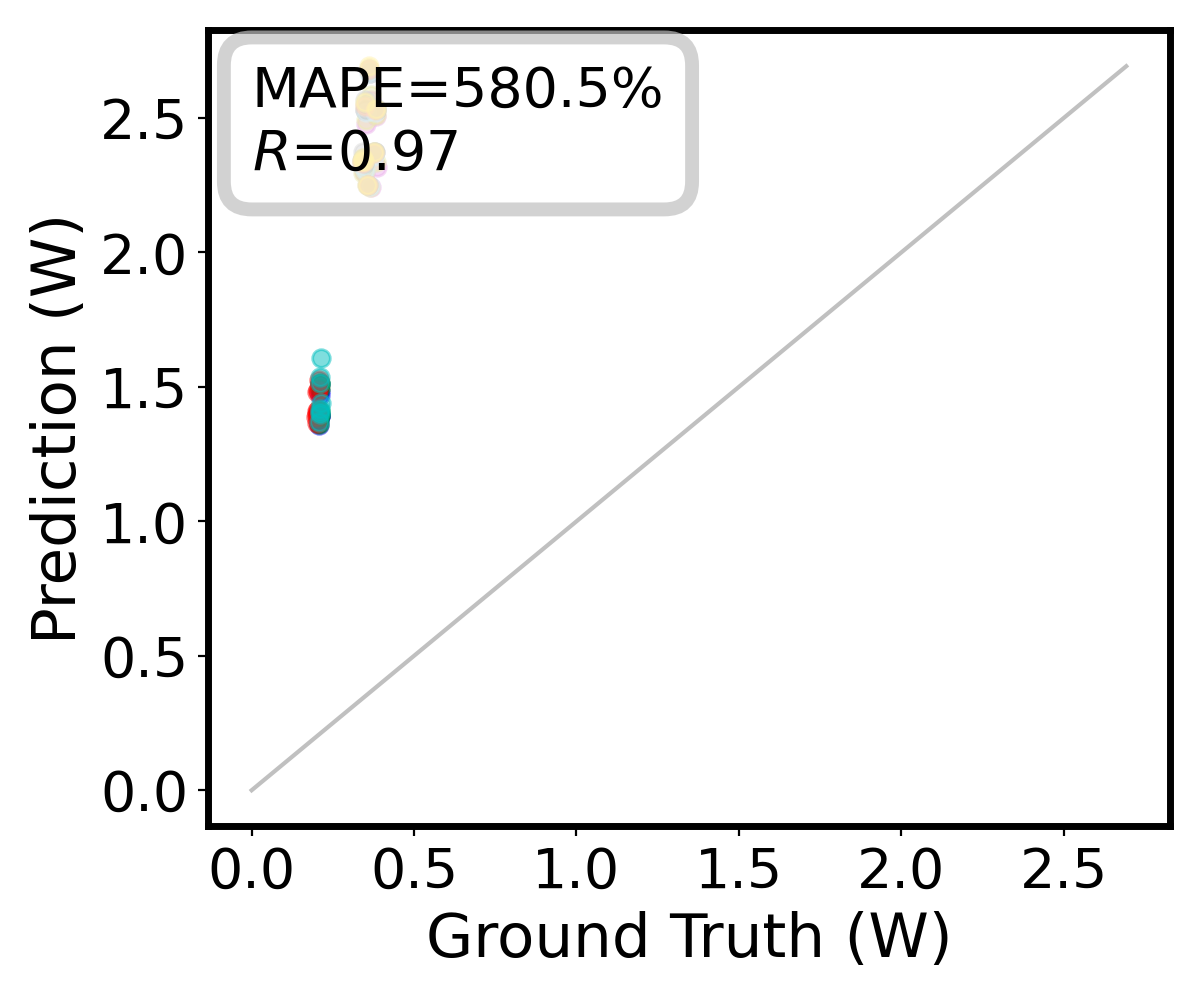}
    \label{techbp}
}
\hspace{-2mm}
\subfigure[ReadyPower]{
    \centering
    \includegraphics[height=0.165\textwidth]{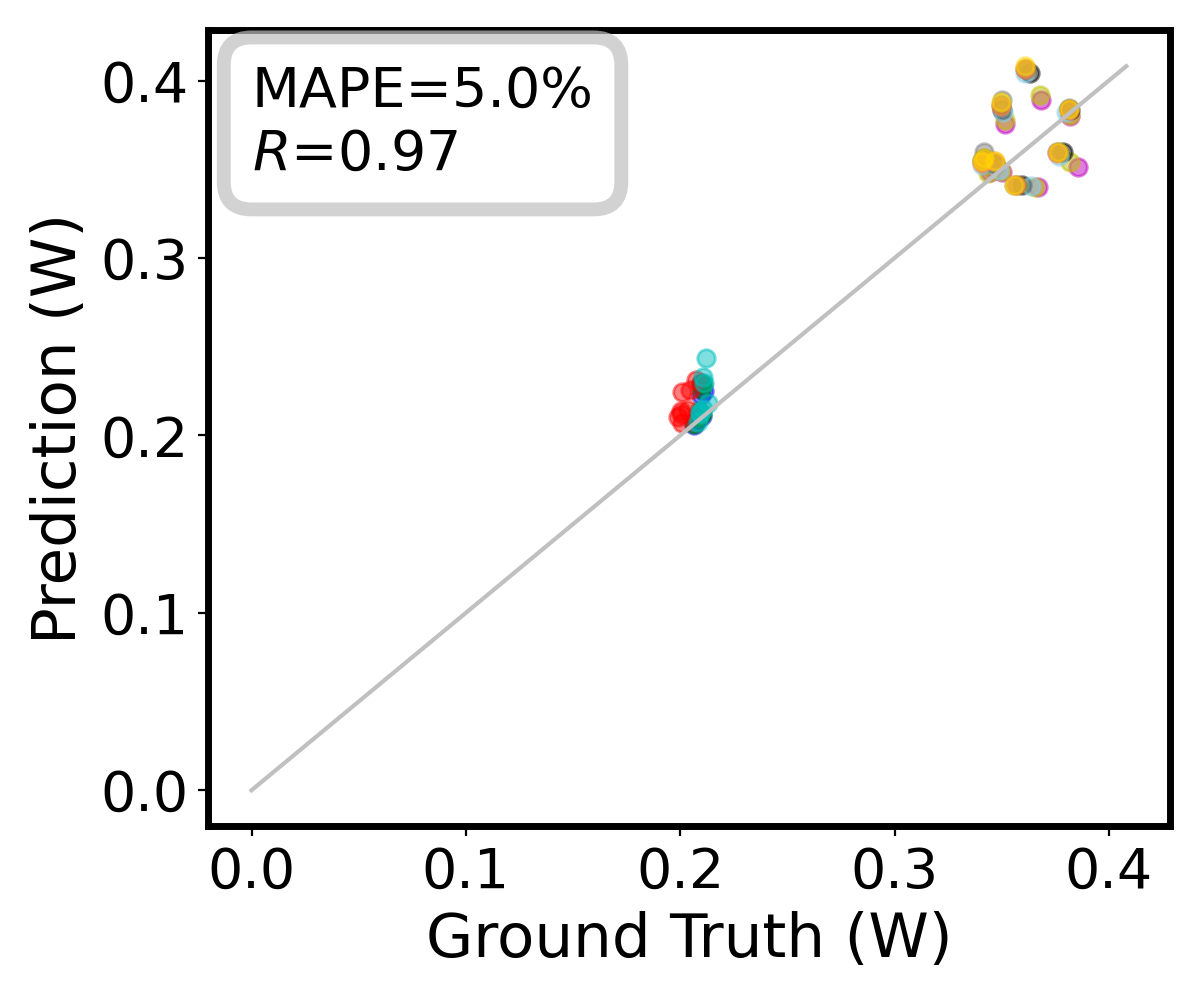}
    \label{whitebp}
}

\vspace{-.1in}
\caption{Illustration of the predictions of ReadyPower and different ablation studies on BOOM branch predictor (BP) as an example.}
\vspace{-.2in}
\label{bpablation}
\end{figure}

\subsection{Ablation Study}

Besides the four baselines discussed above, we also include three methods derived from ReadyPower for ablation studies: ReadyPower-w/o-Arch, ReadyPower-w/o-Impl, and ReadyPower-w/o-Tech, where the architecture-level, implementation-level, and technology-level parameters of ReadyPower are set to the default value in McPAT, respectively. 
Fig.~\ref{ablation} shows the comparison between three derived methods and ReadyPower, with each bar showing the average accuracy value across all training scenarios. It demonstrates that ReadyPower can consistently achieve the lowest MAPE and the highest correlation $R$, indicating that the introduced parameters at all levels are necessary for ReadyPower. Fig.~\ref{bpablation} visualizes the prediction of three derived methods and ReadyPower for the branch predictor (BP), an important processor component, as an example to show the impact of parameters at each level more clearly. 
% in Fig.~\ref{bpablation} to show the impact of parameters at each level more clearly. 
% \looseness=-1

The comparison between ReadyPower and ReadyPower-w/o-Arch validates the effect of the architecture-level parameters in capturing the microarchitecture design discrepancy. The MAPE and correlation gaps between ReadyPower-w/o-Arch and ReadyPower are mainly from the inaccurate prediction of BP and ICache. 
% For example, Fig.~\ref{bpablation}(a) shows the power prediction of BP with ReadyPower-w/o-Arch. 
For the BP, by default, the McPAT regards the BP as an independent component, not depending on the parameters outside the BP. However, the table size of BP usually scales with the FetchWidth in the frontend. Therefore, the ReadyPower-w/o-Arch underestimates the configurations with large FetchWidth, as illustrated in Fig.~\ref{bpablation}(a).

% impl的drop是最大的

The comparison between ReadyPower and ReadyPower-w/o-Impl proves the necessity of implementation-level parameters in capturing the RTL implementation discrepancy. 
The accuracy drop, including high MAPE and low correlation, of ignoring the implementation-level parameters is the most significant among the parameters of the three levels. 
The accuracy drop of ReadyPower-w/o-Impl is mainly due to the inaccurate modeling of the logic scale and table size. 
Fig.~\ref{bpablation}(b) shows that the ReadyPower-w/o-Impl always significantly underestimates the power consumption of BP. It is because of the table size discrepancy of BP between the real processor and the modeled processor in McPAT, where BP is modeled as a simple fixed structure with a small table size.
% The accuracy drop, including high MAPE and low correlation, of ReadyPower-w/o-Impl is mainly due to the inaccurate modeling of the logic scale and table size. 
% Fig.~\ref{bpablation}(b) shows the power prediction of BP with ReadyPower-w/o-Arch as an example. It shows that the ReadyPower-w/o-Impl always significantly underestimates the power consumption of BP. It is because of the table size discrepancy of BP between the real processor and the modeled processor in McPAT, where BP is modeled as a simple fixed structure with a small table size.
% \looseness=-1

The comparison between ReadyPower and ReadyPower-w/o-Tech emphasizes the effect of technology-level parameters in bridging the technology library discrepancy. ReadyPower-w/o-Tech can still retain a comparable correlation $R$, as shown in Fig.~\ref{ablation}, but actually the MAPE drops dramatically. The reason is that the technology library discrepancy can globally affect each part of the processors. 
Fig.~\ref{bpablation}(c) shows the inaccuracy of BP power prediction incurred from the technology library discrepancy, which necessitates the introduction of technology-level parameters.
% Fig.~\ref{bpablation}(c) illustrates the inaccuracy of BP power prediction incurred from the technology library discrepancy. BP is a component dominated by the SRAM array. The comparison between Fig.~\ref{bpablation}(c) and Fig.~\ref{bpablation}(d) shows the inaccurate technology modeling in McPAT, which necessitates the introduction of technology-level parameters.

% Fig.8xxx前面概括

\begin{figure}[!t]
\centering
\vspace{-.2in}
\hspace{-5mm}
\subfigure[40nm]{
    \centering
    \includegraphics[height=0.17\textwidth]{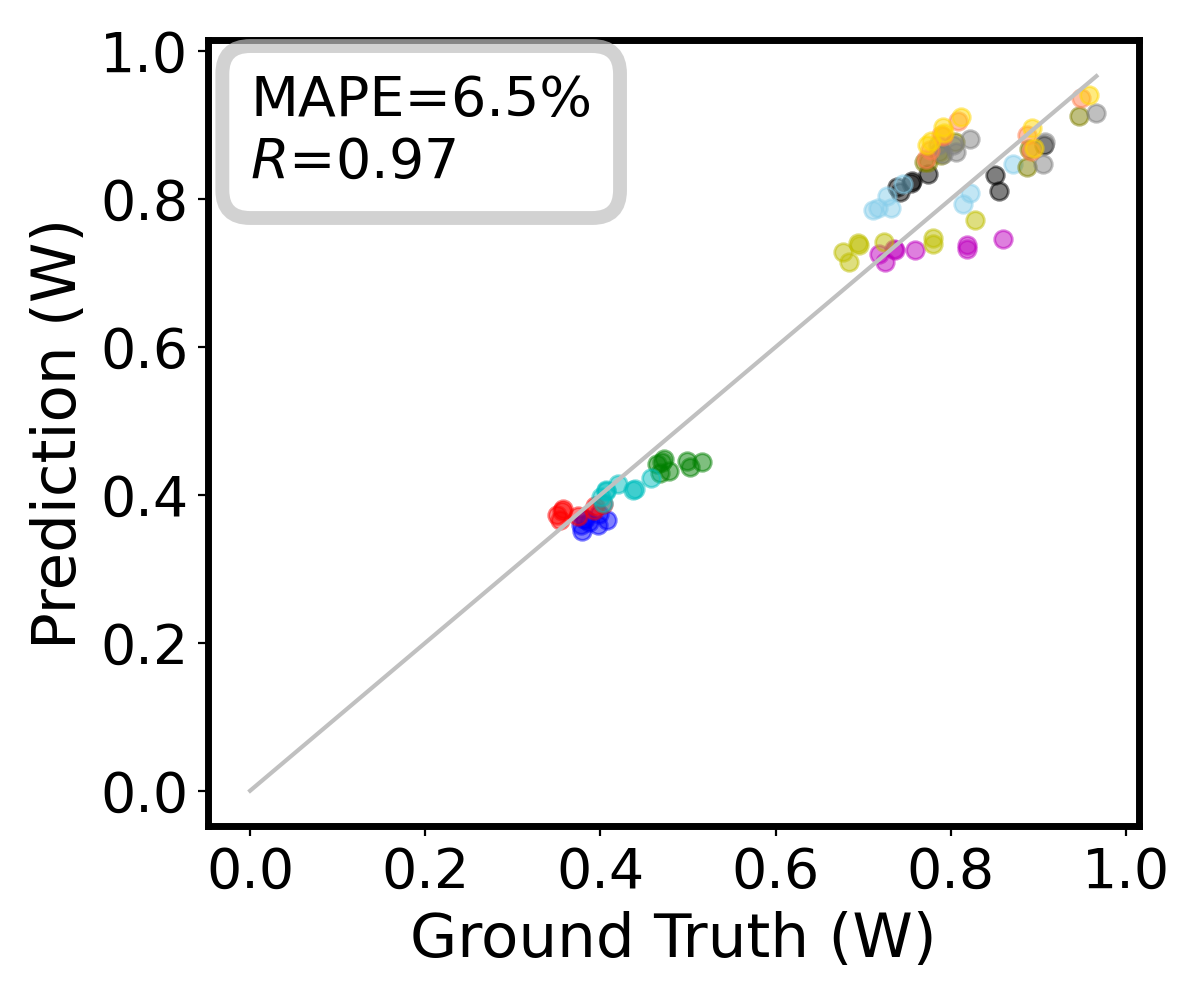}
    \label{40nm}
}
\hspace{-3mm}
\subfigure[28nm]{
    \centering
    \includegraphics[height=0.17\textwidth]{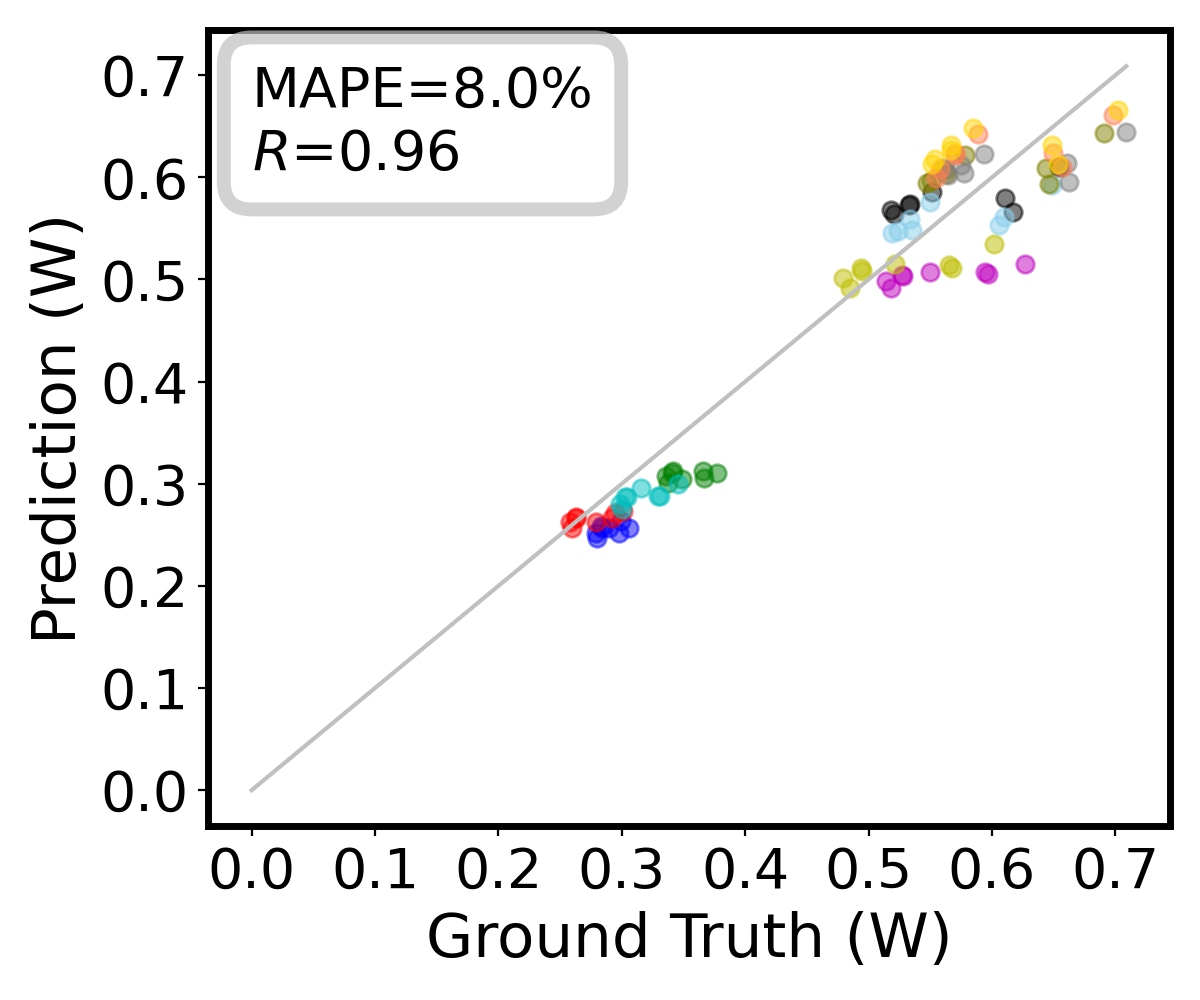}
    \label{28nm}
}

\vspace{-.1in}
\hspace{-5mm}
\subfigure[40nm$\to$28nm]{
    \centering
    \includegraphics[height=0.17\textwidth]{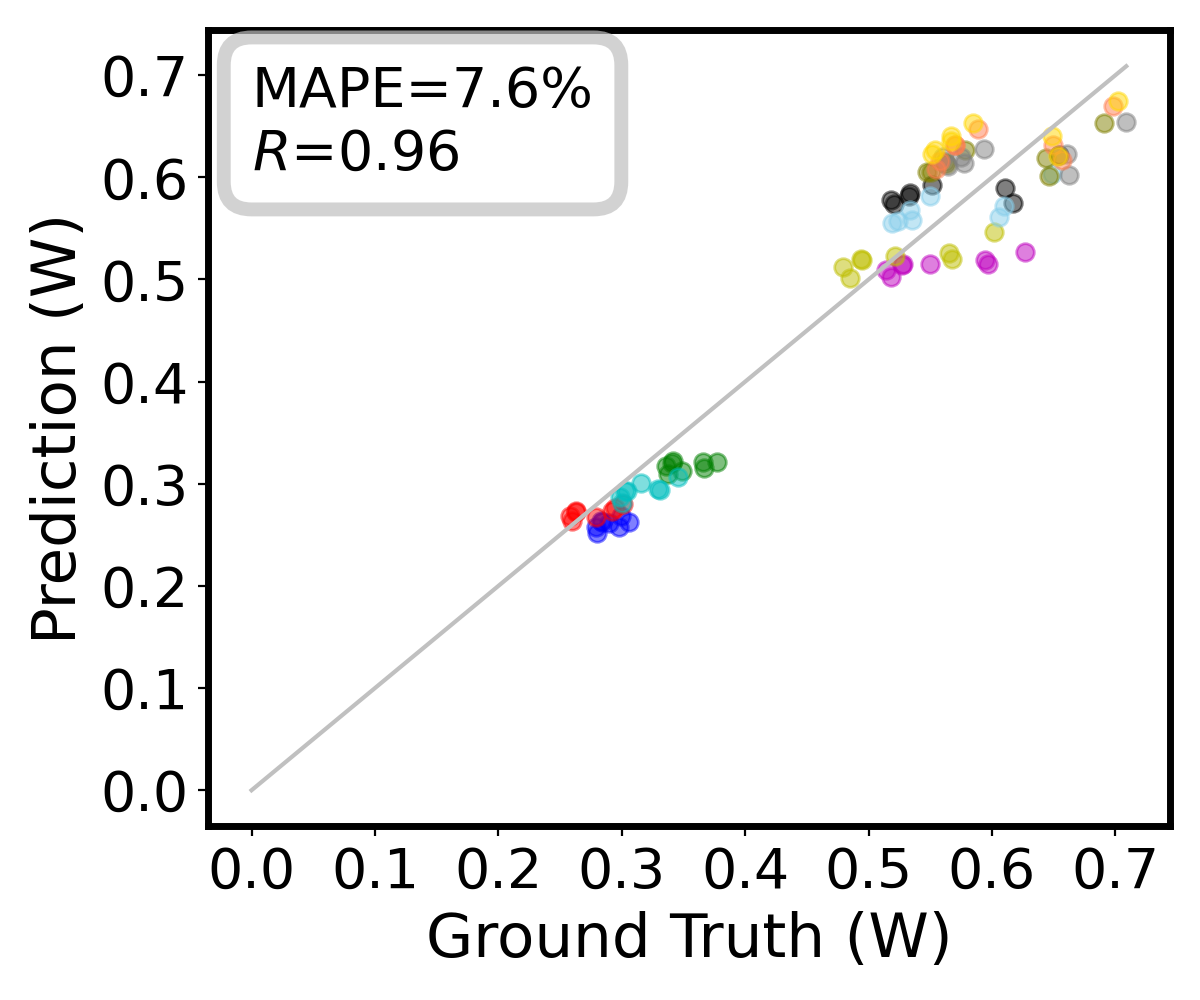}
    \label{40to28}
}
\hspace{-3mm}
\subfigure[28nm$\to$40nm]{
    \centering
    \includegraphics[height=0.17\textwidth]{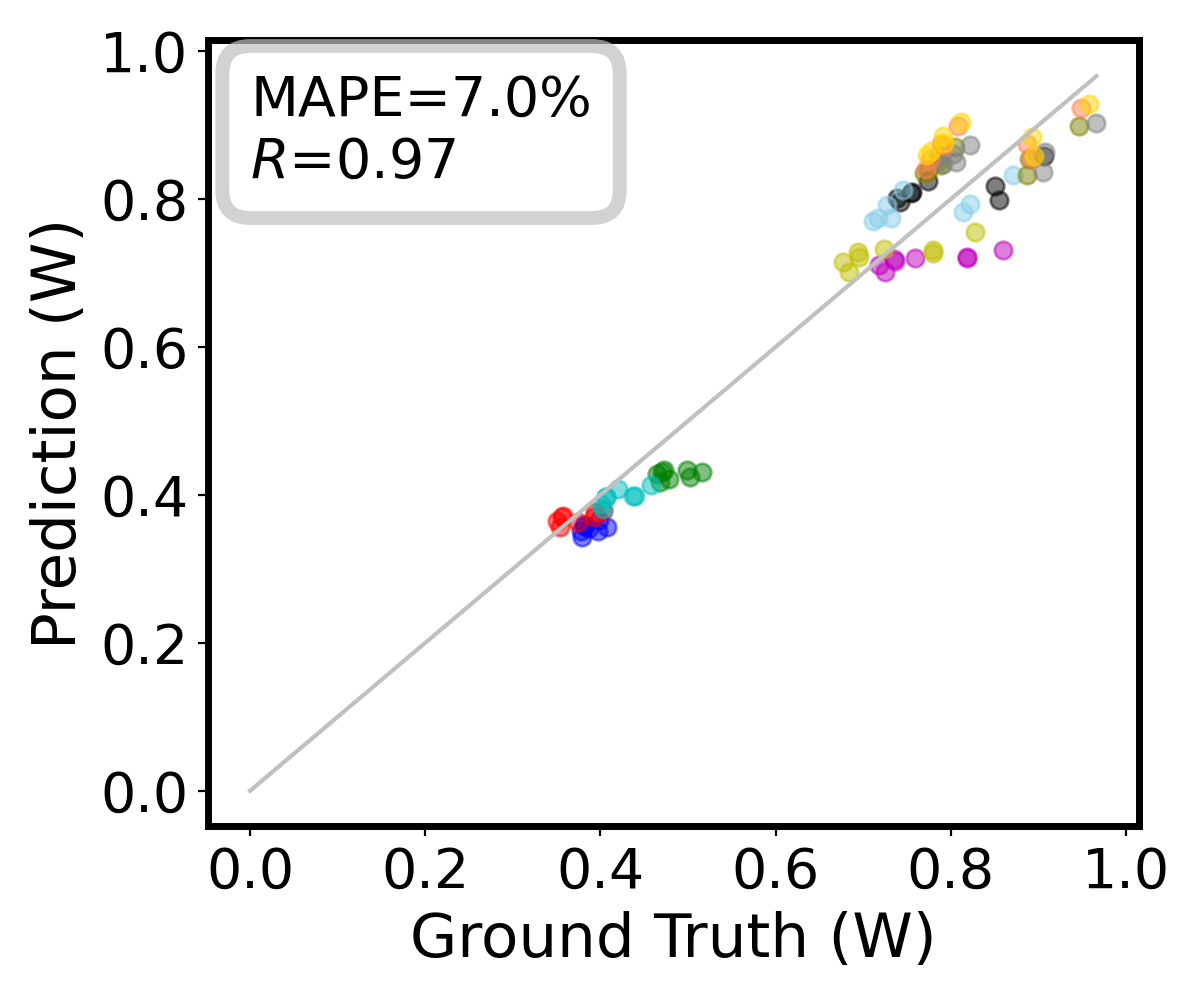}
    \label{28to40}
}

\vspace{-.1in}
\caption{The results of cross-technology-node transfer. (a)(b) Predictions in \emph{Balance} training scenario on source technology node. (c)(d) Predictions in \emph{Balance} training scenario on target technology node.}
\vspace{-.2in}
\label{crossexp}
\end{figure}

\subsection{Cross-Technology-Node Transfer}
A processor design can be implemented with different technology nodes. However, collecting power ground truth for each node requires substantial manpower and runtime overhead. To avoid this repeated cost, transferability across technology nodes is important.

% A processor design can be implemented with different technology nodes. To support different technology nodes without collecting power ground truth repeatedly, which takes significant manpower and runtime overhead, transferring the power model across different technology nodes is important. 
% Because of the manpower and runtime overhead for collecting power ground truth, transferring the power model across different technology nodes is important. 

We evaluate the cross-technology-node transferability of ReadyPower based on the BOOM CPUs implemented with TSMC 28nm and 40nm libraries. When transferring, we re-decide the technology-level parameters with the target technology library while the architecture-level and implementation-level parameters remain unchanged. We perform the transfer from 28nm (source) to 40nm (target) and from 40nm (source) to 28nm (target) with \emph{Balance} training scenario, where \emph{Small} and \emph{Large} training scenarios have a similar trend. Fig.~\ref{crossexp} shows the results, demonstrating the high transferability of ReadyPower.

\section{Conclusion}

In this work, we propose an architecture-level power model named ReadyPower. ReadyPower is ready-for-use by being reliable, interpretable, and handy. 
ReadyPower introduces architecture-level, implementation-level, and technology-level parameters into the widely adopted McPAT analytical model to bridge the discrepancies between the real processor and the modeled processor in McPAT. For each processor architecture, ReadyPower decides the parameters at each level differently. 
Such a ready-for-use framework has high reliability, which is a compelling addition to architects’ toolbox.

\section*{Acknowledgement}
\vspace{-.02in}
This work is supported by National Natural Science Foundation of China (NSFC) 62304192, Hong Kong Research Grants Council (RGC) ECS Grant 26208723, YCRG Grant C6003-24Y, and ACCESS – AI Chip Center for Emerging Smart Systems, supported by the InnoHK initiative of Innovation and Technology Commission of the Hong Kong Special Administrative Region Government.
\vspace{-.05in}

% \newpage
\bibliographystyle{IEEEtran}
\bibliography{references_1, references_2, references_3}
\end{document}